\documentclass[twocolumn]{aastex631}
\usepackage{float}
\usepackage{multirow}
\usepackage{amsmath}
\usepackage{colortbl}
\usepackage{tabularx}
\usepackage{longtable}
\usepackage{booktabs} 
\usepackage{array}
\usepackage{hyperref}

\begin{document}

\title{Luminis Stellarum et Machina: Applications of Machine Learning in Light Curve Analysis}

\correspondingauthor{Nurzhan Ussipov}
\email{Nurzhan.Ussipov@kaznu.edu.kz}

\author[0009-0005-7254-524X]{Almat Akhmetali}
\affiliation{Nonlinear Information Processes Laboratory (NIPL) \\ Department of Electronics and Astrophysics \\
Al-Farabi Kazakh National University, 050040 Almaty, Kazakhstan}

\author[0009-0002-5435-9740]{Alisher Zhunuskanov}
\affiliation{Nonlinear Information Processes Laboratory (NIPL) \\ Department of Electronics and Astrophysics \\
Al-Farabi Kazakh National University, 050040 Almaty, Kazakhstan}

\author[0009-0001-8784-4470]{Aknur Sakan}
\affiliation{Nonlinear Information Processes Laboratory (NIPL) \\ Department of Electronics and Astrophysics \\
Al-Farabi Kazakh National University, 050040 Almaty, Kazakhstan}

\author[0009-0006-8505-7277]{Marat Zaidyn}
\affiliation{Nonlinear Information Processes Laboratory (NIPL) \\ Department of Electronics and Astrophysics \\
Al-Farabi Kazakh National University, 050040 Almaty, Kazakhstan}

\author[0000-0002-2389-2262]{Timur Namazbayev}
\affiliation{Nonlinear Information Processes Laboratory (NIPL) \\ Department of Electronics and Astrophysics \\
Al-Farabi Kazakh National University, 050040 Almaty, Kazakhstan}

\author[0000-0002-7326-9196]{Dana Turlykozhayeva}
\affiliation{Nonlinear Information Processes Laboratory (NIPL) \\ Department of Electronics and Astrophysics \\
Al-Farabi Kazakh National University, 050040 Almaty, Kazakhstan}

\author[0000-0002-2512-3280]{Nurzhan Ussipov}
\affiliation{Nonlinear Information Processes Laboratory (NIPL) \\ Department of Electronics and Astrophysics \\
Al-Farabi Kazakh National University, 050040 Almaty, Kazakhstan}

\begin{abstract}

The rapid advancement of observational capabilities in astronomy has led to an exponential growth in the volume of light curve (LC) data, creating both opportunities and challenges for time-domain astronomy. Traditional analytical methods often struggle to fully extract the scientific value of these large and complex datasets. Machine learning (ML) algorithms are increasingly used for LC analysis, enabling classification, prediction, pattern discovery and anomaly detection. However, research in this area remains fragmented, with no comprehensive synthesis of how ML methods address the specific challenges of LC data. Key difficulties include class imbalance, noisy or sparse measurements, effective feature extraction, and limited interpretability of models. This lack of a unified overview makes it difficult for researchers to identify suitable approaches or recognize unresolved problems that require methodological advances. To address this gap, this survey systematically reviews ML techniques applied to LC analysis, outlining their principles and applications in tasks such as exoplanet detection, variable star classification and supernova identification. By clarifying the current state of the field and highlighting open challenges, this work provides guidance for future research and supports the effective integration of ML into astronomical big data studies.

\end{abstract}


\section{Introduction} \label{sec:Introduction }

Time domain astronomy (TDA) is a rapidly evolving field, driven by the continual discovery of new phenomena and the expansion of observational capabilities~\cite{ball1, pesenson2}. This field investigates time-varying characteristics of celestial objects using data from diverse messengers, including gravitational waves, neutrinos, and electromagnetic radiation across various photon-energy bands~\cite{vaughan3}. With the advent of advanced wide-field, multi-epoch sky surveys such as the Sloan Digital Sky Survey (SDSS)~\cite{york4}, Zwicky Transient Facility (ZTF)~\cite{graham5}, Panoramic Survey Telescope and Rapid Response System (Pan-STARRS)~\cite{kaiser6}, and the upcoming Vera C. Rubin Observatory (formerly LSST)~\cite{ivezic7}, the volume of transient discoveries has grown exponentially. For instance, ZTF generates alerts for over 100,000 events each night, while LSST is projected to surpass this by an order of magnitude. These alerts, which indicate significant flux density changes or new spatial positions, represent invaluable data streams for studying transient phenomena such as supernovae, pulsars, and gamma-ray bursts. TDA focuses on analyzing time-series data to construct LCs, which depict the variation in brightness of celestial objects over time and are fundamental to understanding the physical processes driving these phenomena.

However, harnessing these vast data streams requires addressing significant challenges. It is important for astronomers to take into account the possible uncertainties and biases that may arise from observational methods as well as from data analysis procedures. Observational uncertainties stem from limitations in instrumentation and data collection. Analytical uncertainties, on the other hand, emerge from the complexities of modeling physical processes and the numerical methods used to approximate them. Mitigating biases requires developing comprehensive models that incorporate diagnostics from multiple perspectives. This necessitates detailed multi-physics simulations that combine data from various sources and rely on cutting-edge computational capabilities. As a result, TDA not only facilitates the study of astrophysical phenomena but also serves as a bridge to advancing fundamental physics, making it a cornerstone of modern astronomical research.

ML has become an essential tool for automating data classification, archiving, and retrieval~\cite{fluke9, baron10}. Unlike traditional feature extraction methods, which require domain expertise, deep learning (DL) automatically learns data representations. As a subset of ML, DL has gained widespread popularity and has been successfully applied in various domains, including speech recognition, image processing, and natural language processing~\cite{nassif11, deng12, rodellar13, mishra14, ussipov15, ussipov16, otter21}. In astronomy, ML has emerged as a powerful tool for automating the classification and analysis of LCs, enabling researchers to extract insights from vast datasets.

Despite its growing adoption, several persistent challenges limit the effectiveness of ML for LC analysis. Class imbalance occurs because some astronomical sources, such as rare transients, are vastly underrepresented compared to more common classes, which can bias models toward majority classes and reduce sensitivity to rare events. Noise arises from observational uncertainties, instrumental limitations, or environmental effects, often degrading model performance and leading to spurious detections. Interpretability is another major challenge, as many ML and DL methods function as “black boxes,” making it difficult to relate predictions to physical processes or diagnose errors. In addition, biases and uncertainties can stem from selection effects in survey design, incomplete or mislabeled training data, and systematic measurement errors; these issues can propagate through the ML pipeline and affect generalizability. Addressing these challenges is crucial for building models that not only produce accurate predictions but also offer trustworthy scientific insights.

This article provides a comprehensive overview of ML algorithms in LC analysis, covering applications such as exoplanet detection, variable star classification, and supernova identification. By presenting a structured examination of ML techniques in this domain, the review aims to foster a deeper understanding of current methodologies and identify chanllenges that should be adressed in the near feature. 

The rest of the paper is organized as follows. The methodology is presented in Section~\ref{sec:Methodology}. Section~\ref{sec:Photometric} provides an overview of major photometric data surveys relevant to LC analysis. Section 4 
introduces fundamental ML concepts and architectures. Section~\ref{sec:Application} explores key applications of ML techniques to LC analysis. Section~\ref{sec:Challenges} outlines current challenges and open research questions in the field. Finally, Section~\ref{sec:Conclusions} concludes with a summary and outlook.

\section{Review Methodology}\label{sec:Methodology}  

\begin{figure*}[htbp]
    \centering
   \includegraphics[width=0.85\textwidth]{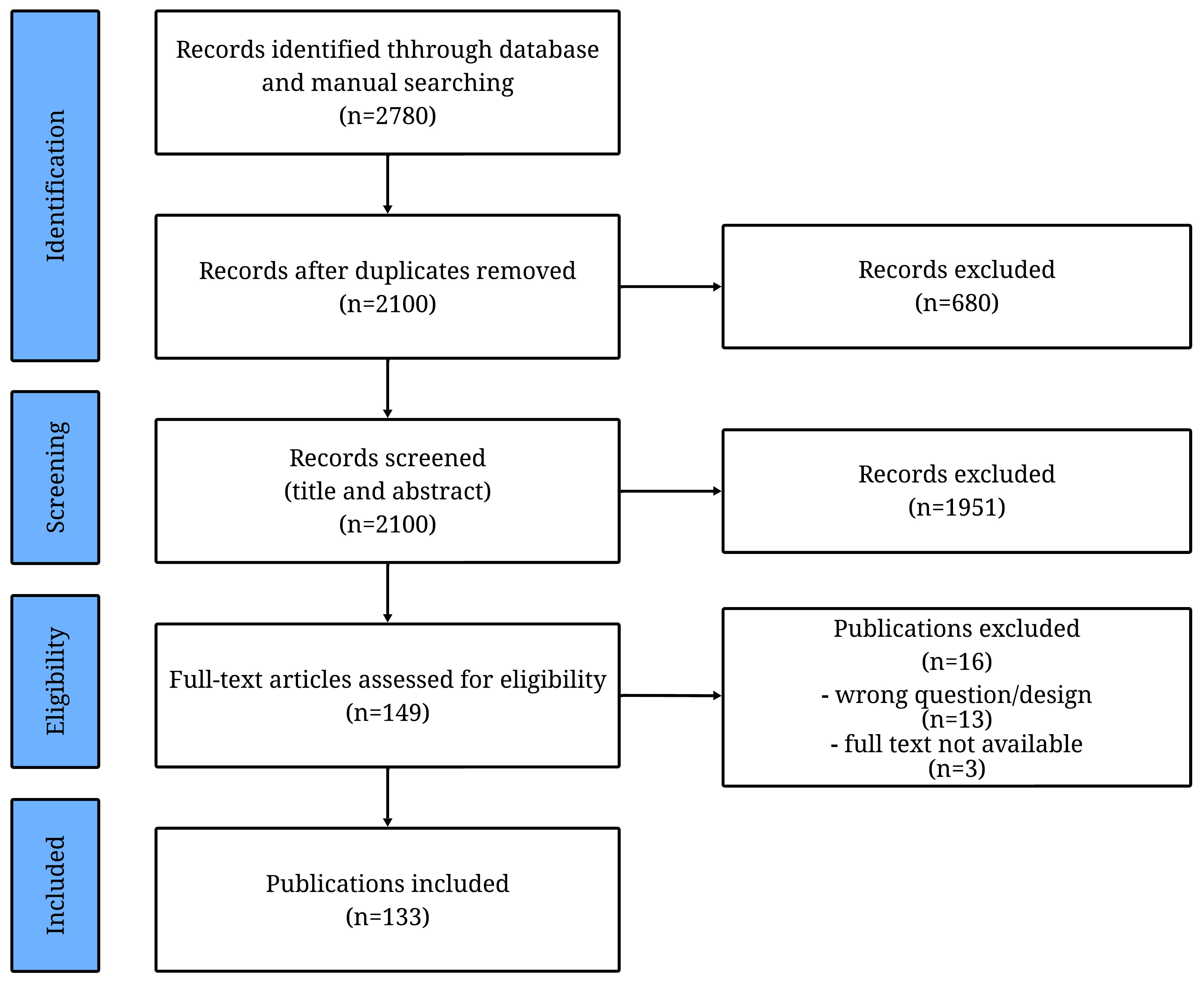}
    \caption{PRISMA flowchart of review methodology.}
    \label{fig:PRISMA}
\end{figure*}

To compile this review, we followed the Preferred Reporting Items for Systematic Reviews and Meta-Analyses (PRISMA) guidelines \cite{moher2009preferred}. The overall review methodology is illustrated in Figure~\ref{fig:PRISMA}. A comprehensive search was conducted across major scientific databases, including IEEE Xplore, SpringerLink, Elsevier, Web of Science, and Google Scholar, using Boolean combinations of keywords such as “light curve classification”, “light curve machine learning”, “variable star classification”, “supernova classification”, and “exoplanet detection”. The initial search identified 2,780 records, of which 680 duplicates were removed. Title and abstract screening excluded an additional 1,951 records, followed by the removal of 16 records after full-text eligibility assessment. Ultimately, 133 studies were included in this review.

The review covers publications from 2005 to 2025. Studies were included if they (1) were peer-reviewed, (2) applied ML methods to real or simulated LC data, and (3) were relevant to tasks such as variable star classification, exoplanet detection, or transient identification. Non-English papers, duplicate entries, and studies lacking experimental validation or methodological clarity were excluded.

\section{Photometric Survey Datasets}\label{sec:Photometric}  

\begin{figure}[h]
    \centering
    \includegraphics[width=1\linewidth]{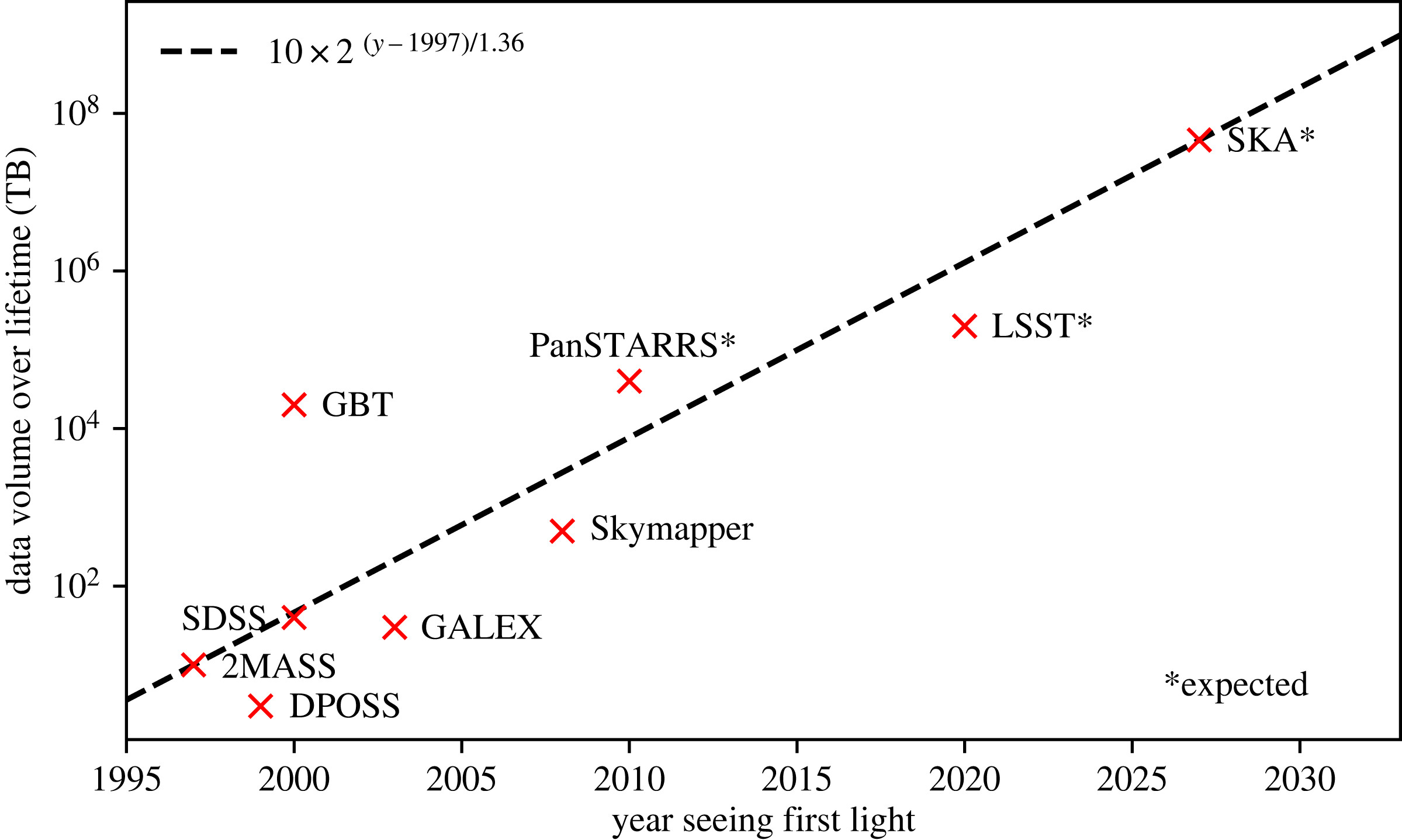}
    \caption{The total data volume produced by various astronomical surveys throughout their lifetimes. The data reveals that the volume of astronomical survey data doubles approximately every 16 months. The data and figure are sourced from~\cite{kremer52, smith53}.}
    \label{fig:Big Data}
\end{figure}

Over the past few decades, large-scale photometric surveys have transformed time-domain astronomy, providing vast, multi-wavelength datasets that are readily accessible through online databases. These surveys have been essential in detecting and classifying variable stars, discovering exoplanets, and identifying transient astronomical events. As shown in Figure~\ref{fig:Big Data}, the data volume from astronomical surveys has grown exponentially, approximately doubling every 16 months. The combination of different survey datasets enables comprehensive studies of stellar variability across various timescales and wavelengths, driving both traditional astrophysical research and modern ML applications. This rapid data growth has created a strong demand for automated and scalable analysis methods, especially those based on ML.

Understanding the properties of these surveys is essential for ML applications, since survey-specific factors such as cadence, photometric precision, time coverage, and instrumental effects directly affect the structure and quality of LC data. In addition, biases and uncertainties introduced during survey design or data processing can influence ML model performance, generalization, and reliability. Summarizing major photometric surveys provides important context for selecting appropriate ML techniques and for understanding the practical limitations of real-world astronomical data.

Photometric surveys can be broadly categorized into ground-based and space-based missions. Ground-based surveys provide wide-field and high-cadence monitoring, often covering extensive regions of the sky over long time periods. In contrast, space-based missions offer high-precision, uninterrupted observations unaffected by atmospheric distortions, enabling the detection of minute variations in stellar brightness.  
Table~\ref{table1} provides an overview of major ground-based and space-based photometric surveys, including their active years, observed bands, sky coverage, and references.  

\begin{longtable*}{|m{6cm}|m{2cm}|m{2cm}|m{2cm}|m{4cm}|}
    \caption{Overview of major photometric surveys, including their active years, observed bands, and sky coverage.}
    \label{table1} \\
    \hline
    \textbf{Surveys} & \textbf{Active years} & \textbf{Bands} & \textbf{Area} & \textbf{Reference} \\ \hline 
    \endfirsthead
    
    \multicolumn{5}{c}{{\tablename\ \thetable{} - Continued}} \\
    \hline
    \textbf{Surveys} & \textbf{Active years} & \textbf{Bands} & \textbf{Area} & \textbf{Reference} \\ \hline 
    \endhead
    
    \hline
    \endfoot
    
    \hline
    \endlastfoot

    Hipparcos / Tycho & 1989-1993 & $B_T$, $V_T$ & MW &~\cite{perryman23, hog24} \\ \hline
    MAssive Compact Halo Objects (MACHO) & 1992-1999 & V, R & MW, LMC, SMC &~\cite{alcock25} \\ \hline
    Optical Gravitational Lensing Experiment (OGLE) & 1992-present & V, I & MW, LMC, SMC &~\cite{udalski26} \\ \hline
    All Sky Automated Survey (ASAS) & 1997-present & V, I & All sky &~\cite{pojmanski27} \\ \hline
    Two Micron All-Sky Survey (2MASS) & 1997-2001 & J, H, $K_s$ & All sky &~\cite{skrutskie28} \\ \hline
    Lincoln Near-Earth Asteroid Research (LINEAR) & 1998-2015 & Unfiltered & Wide field &~\cite{stokes29} \\ \hline
    Robotic Optical Transient Search Experiment (ROTSE) & 1998-present & Unfiltered & All sky &~\cite{akerlof30} \\ \hline
    Sloan Digital Sky Survey (SDSS) & 1998-present & u, g, r, i, z & Stripe 82 &~\cite{york4} \\ \hline
    Northern Sky Variability Survey (NSVS) & 1999-2004 & Unfiltered & Northern sky &~\cite{wozniak31} \\ \hline
    Catalina Real-Time Survey (CRTS) & 2003-present & V & All sky &~\cite{drake32} \\ \hline
    Hungarian Automated Telescope Network (HATnet) & 2003-present & r & Wide field &~\cite{bakos33} \\ \hline
    Wide Angle Search for Planets (WASP / SuperWASP) & 2004-present & Optical & All sky &~\cite{pollacco34} \\ \hline
    UKIRT Infrared Deep Sky Survey (UKIDSS) & 2005-2014 & Z, Y, J, H, K & Wide field &~\cite{lawrence35} \\ \hline
    Convection, Rotation and Planetary Transits (CoRoT) & 2006-2013 & Unfiltered & MW &~\cite{barge36} \\ \hline
    Kepler mission & 2009-2018 & Unfiltered & MW &~\cite{kochі37} \\ \hline
    Wide field Infrared Survey Explorer (WISE) & 2009-present & $W_1$, $W_2$, $W_3$, $W_4$ & Wide field &~\cite{wright38} \\ \hline
    VISTA Variables in the Vía Láctea (VVV) & 2010-2016 & Z, Y, J, H, $K_s$ & Wide field &~\cite{minniti39} \\ \hline
    High Cadence Transit Survey (HiTS) & 2013-2015 & u, g, r, i & Wide field &~\cite{forster40} \\ \hline
    \textit{Gaia} & 2013-present & BP, RP  & All sky &~\cite{prusti41} \\ \hline
    All Sky Automated Survey for SuperNovae (ASAS-SN) & 2014-present & V, g & All sky &~\cite{kochanek42} \\ \hline
    Next Generation Transit Survey (NGTS) & 2015-present & I & Wide field &~\cite{wheatley43} \\ \hline
    Panoramic Survey Telescope and Rapid Response System (Pan-STARRS) & 2010-present & g, r, i, z, y & Wide field &~\cite{magnier44} \\ \hline
    Transiting Exoplanet Survey Satellite (TESS) & 2018-present & TESS-band & All sky &~\cite{ricker45} \\ \hline
    Zwicky Transient Facility (ZTF) & 2018-present & g, r, i & Northern sky &~\cite{bellm46} \\ \hline
    James Webb Space Telescope (JWST) & 2021-present & IR  & Targeted fields &~\cite{gardner47} \\ \hline
    Wide Field Survey Telescope (WFST) & 2023-present & u, g, r, i, z, w & Northern sky &~\cite{lou48} \\ \hline
    Vera C. Rubin Observatory (formerly LSST) & 2025 (expected) & u, g, r, i, z, y & Wide field &~\cite{ivezic7} \\ \hline
    PLAnetary Transits and Oscillations of Stars (PLATO) & 2026 (expected) & B, V, R, I & Wide field &~\cite{rauer49} \\ \hline
    Ultraviolet Transient Astronomy Satellite (ULTRASAT) & 2027 (expected) & NUV  & All sky &~\cite{shvartzvald50} \\ \hline
    Roman Space Telescope (formerly WFIRST) & 2027 (expected) & IR & Wide field &~\cite{spergel51} \\ \hline
    
\end{longtable*}

\section{Machine learning fundamentals}
\label{sec:Machine}
To demonstrate the applications of ML in LC analysis, this section provides a concise overview of core ML concepts.

\begin{figure*}[htbp]
    \centering
   \includegraphics[width=0.9\textwidth]{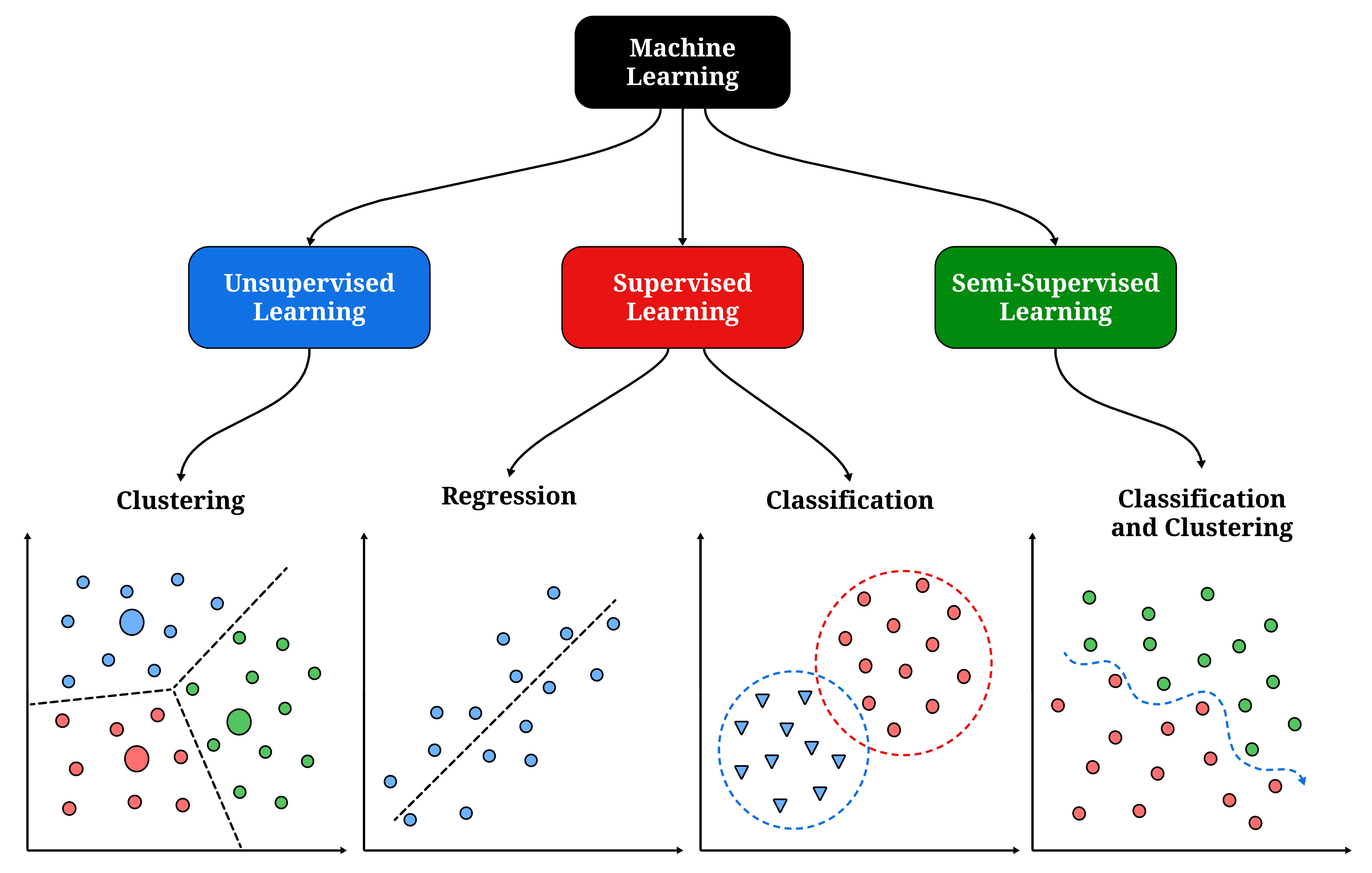}
    \caption{Overview of main ML approaches.}
    \label{fig:ML_taxonomy}
\end{figure*}

\begin{figure*}[htbp]
    \centering
   \includegraphics[width=0.9\textwidth]{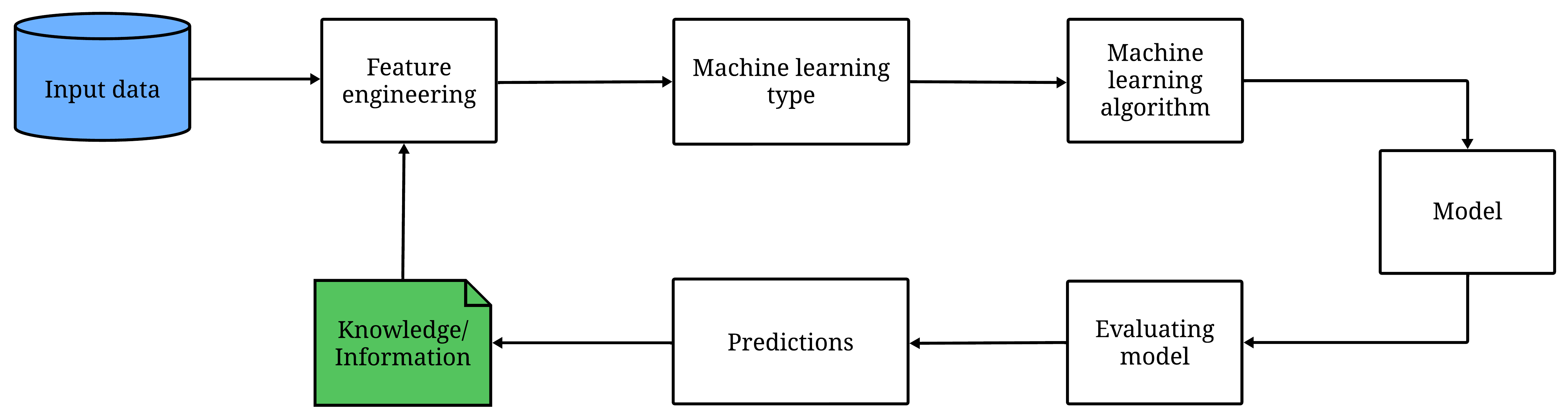}
    \caption{Basic process of ML. Cylinder represents database, rectangles indicate data processing steps, and the rectangle with a folded corner denotes knowledge extraction.}
    \label{fig:ML}
\end{figure*}

ML enables systems to autonomously learn patterns from data and improve decision-making through experience, mirroring aspects of human cognition. Unlike traditional astronomical programming, which relies on explicit physical rules, ML algorithms derive implicit relationships directly from observational data. This data-driven approach offers flexibility for solving complex, nonlinear problems that defy conventional analytical methods~\cite{rodriguez54, kembhavi55, sen56}.

ML algorithms can be broadly categorized into supervised and unsupervised methods, often referred to as predictive and descriptive, respectively. These approaches may also be combined to form semi-supervised methods. Figure~\ref{fig:ML_taxonomy} provides a visual taxonomy of these approaches. Supervised algorithms learn mappings between input features and predefined target variables using labeled training data curated by domain experts~\cite{zucker79, delli80, krone81, mahabal82, d83, norris84}. Unsupervised methods, conversely, autonomously identify hidden structures or relationships within unlabeled datasets. These techniques are typically divided into three subcategories: clustering (grouping similar data points), dimensionality reduction (extracting salient features), and anomaly detection (identifying outliers)~\cite{hocking96, nun97, gianniotis98, baron99, reis100, reis101}. Anomaly detection holds particular promise for astronomical research, as it enables discovery of rare or unexpected phenomena within existing observational datasets.

Another major category is reinforcement learning, which is based on learning through interaction with an environment to maximize cumulative rewards. However, it is excluded from the scope of this review, as its application to LC analysis remains limited due to the lack of interactive, feedback-driven settings in typical observational astronomy workflows.

As an interdisciplinary field, ML integrates principles from statistics, optimization theory, and information science. Its primary focus lies in developing adaptive systems that simulate human learning processes to iteratively acquire knowledge, refine skills, and optimize performance. Figure~\ref{fig:ML} illustrates the iterative workflow of a typical ML system. Knowledge extraction from data can be effectively achieved using established methodologies such as the Knowledge Discovery in Databases (KDD) process~\cite{fayyad101b} and the Cross-Industry Standard Process for Data Mining (CRISP-DM)~\cite{wirth101c}.

\subsection{Supervised Learning}\label{subsec:Supervised}  

\begin{figure*}[htbp]
    \centering
   \includegraphics[width=0.9\textwidth]{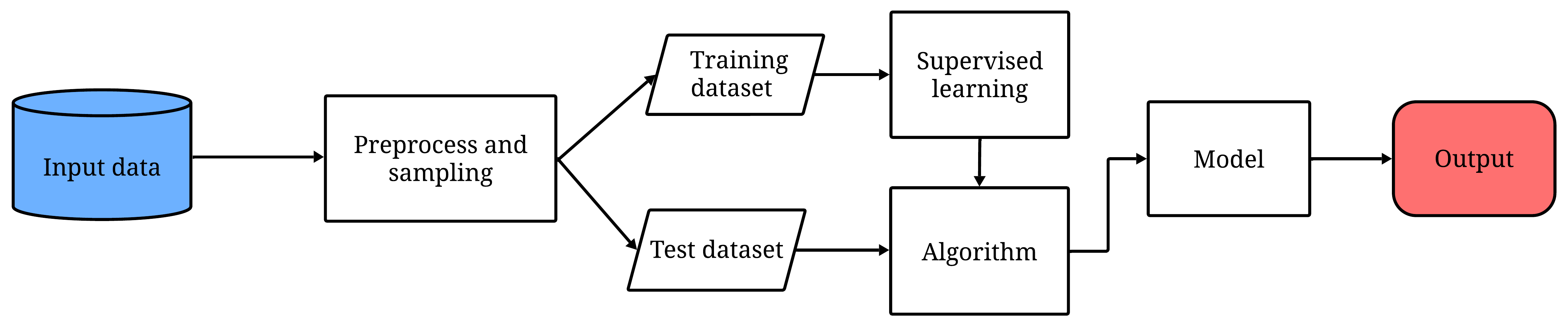}
    \caption{Basic process of SL. Cylinder represents database, rectangles indicate data processing steps, parallelograms denote datasets, and the rounded rectangle shows the final output.}
    \label{fig:SL}
\end{figure*}

Supervised learning (SL) operates under guidance, where labeled data provides the necessary supervision for model training. In this framework, class labels or continuous target values act as references, enabling the model to learn mappings between inputs and outputs. The basic flow of SL is shown in Figure~\ref{fig:SL}. SL problems are typically divided into classification and regression. Classification involves predicting discrete class labels (e.g., variable stars based on their LC), while regression involves predicting continuous values (e.g., estimating a star’s period or temperature from its LC). Both approaches have been applied to LC analysis: classification is widely used for star type identification, supernova classification and exoplanet detection~\cite{akhmetali244, mahabal211, valizadegan178, pearson148,mitra2023exploring, mitra2024probing,abylkairov2025evaluating, bissekenov2024}, whereas regression is used for parameter estimation and LC detrending~\cite{hinners212, thieler101d, aigrain101e, paunzen101f, carter101g}. 

These algorithms optimize their performance by minimizing a cost function, which quantifies the discrepancy between predicted and actual values. The greater the deviation, the more challenging it becomes to achieve accurate predictions. Effective learning depends on a well-curated dataset with precise class labels, as a larger and higher-quality training set facilitates smoother optimization and improves model accuracy.

\begin{figure*}[htbp]
    \centering
   \includegraphics[width=0.9\textwidth]{Figures/SL_algorithms.pdf}
    \caption{Overview of SL algorithms. (a) RF: An ensemble of DTs with majority voting or averaging. (b) MLP: A feedforward neural network with input, hidden, and output layers. (c) RNN: Processes sequential data using recurrent connections (denoted as red lines) between hidden layers.  (d) LSTM: A specialized RNN with memory cells to capture long-term dependencies, utilizing recurrent connections (denoted as red lines) for information flow.  (e) SVM: Identifies an optimal hyperplane with maximum margin for class separation.  (f) KNN: Classifies data based on the majority vote of the $k$ nearest neighbors. (g) CNN: Extracts hierarchical features using convolutional and pooling layers, followed by fully connected layers for classification.}
    \label{fig:SL_algorithms}
\end{figure*}

Decision Trees (DTs) are hierarchical, non-parametric models that recursively partition the feature space by applying conditions on features at internal nodes~\cite{quinlan103}. The model is defined by this tree structure, with nodes, branches, and leaves representing features, decisions, and outcomes, respectively.
The learning algorithm trains the tree by selecting splits that minimize impurity measures such as entropy for classification or mean squared error (MSE) for regression:
\[
\text{Entropy}(R) = -\sum_{k=1}^K p_k \log_2 p_k,
\]

\[
\text{MSE}(R) = \frac{1}{|R|} \sum_{i \in R} (y_i - \bar{y}_R)^2,
\]
where \( p_k \) is the probability of class \( k \) in region \( R \), and \( \bar{y}_R \) is the mean of the target values in \( R \). The algorithm also includes strategies to mitigate overfitting and bias, such as pruning or limiting tree depth, and handles class imbalance using cost-sensitive splitting or balanced sampling. DTs naturally quantify uncertainty via the class probability distribution at leaf nodes. DTs are used for star–galaxy separation~\cite{vasconcellos103a}, detecting active objects~\cite{zhao103b}, redshift estimation~\cite{suchkov103c}, and classifying stars, galaxies, and AGNs~\cite{golob103d}.

Random Forests (RFs) are ensembles of \( B \) DTs, which together define the model~\cite{breiman104}. Each tree partitions the feature space into regions corresponding to its leaf nodes, where a region \(R\) contains all samples reaching that leaf. The learning algorithm trains each tree on bootstrap samples with random feature subsets. Predictions are then aggregated from all trees via majority voting for classification or averaging for regression: 
\[
\hat{y} = \text{majority vote} \left( \{ f_b(x) \}_{b=1}^B \right) \quad \text{or} \quad \hat{y} = \frac{1}{B} \sum_{b=1}^B f_b(x),
\]
Randomization reduces correlation among trees and enhances robustness. Uncertainty can be estimated using out-of-bag (OOB) predictions as an internal validation measure. Random feature selection mitigates bias and overfitting, while class imbalance is handled through balanced class weights. Figure~\ref{fig:SL_algorithms}(a) illustrates a typical RF architecture. RFs are widely used for supervised tasks such as variable star or supernova classification~\cite{jenkins156, kim207, gomez257}, and can also be applied in unsupervised settings to estimate object similarity~\cite{reis100, reis101}.

Artificial Neural Networks (ANNs) are nonlinear, adaptive computational models composed of interconnected units called neurons~\cite{haykin106}. The model is defined by the network architecture, specifying layers, neurons, and connectivity. Each neuron computes an output:
\[
y = \phi(w^T x + b),
\]
where \(\phi\) is an activation function, \(w\) is a vector of connection weights, \(x\) is the input vector, and \(b\) is a bias term. The learning algorithm, such as backpropagation, adjusts the weights to minimize prediction error. Inspired by biological neural networks, ANNs simulate learning and decision-making by adapting connections based on input data, making them highly effective for complex pattern recognition and predictive modeling in high-dimensional domains. Class imbalance is addressed using weighted loss functions, while uncertainty can be approximated via dropout or Bayesian layers. ANNs are used for classification of galaxy spectra~\cite{folkes106a}, modeling Spectral Energy Distribution (SED) of galaxies~\cite{silva106b}, probing the opacity of Universe to cosmic gamma rays~\cite{singh106c}, and prediction of the atmospheric fundamental parameters from stellar spectra~\cite{azzam106d}.

Multilayer Perceptrons (MLPs) are feedforward ANNs composed of an input layer, one or more fully connected hidden layers, and an output layer. The model is defined by the architecture specifying the number of layers, neurons per layer, and activation functions. Each hidden layer computes:
\[
\mathbf{h}_l = \sigma(\mathbf{W}_l \mathbf{h}_{l-1} + \mathbf{b}_l),
\]
where \(\sigma\) is an activation function, typically the rectified linear unit (ReLU)~\cite{pal107}, \(\mathbf{W}_{l}\) and \(\mathbf{b}_{l}\) are the weight matrix and bias vector for layer \(l\), and \(\mathbf{h}_{0} = \mathbf{x}\) is the input vector. The learning algorithm, such as backpropagation, adjusts the weights to minimize loss functions such as cross-entropy (classification) or MSE (regression). Uncertainty can be modeled using dropout or Bayesian layers. Regularization techniques such as L2 weight decay reduce bias and prevent overfitting. Class imbalance is typically addressed by applying class-weighted loss functions or oversampling minority classes. Figure~\ref{fig:SL_algorithms}(b) illustrates a typical MLP architecture. MLPs are widely used in astronomy for tasks such as star–galaxy classification and distinguishing galaxies, stars, and quasars~\cite{andreon107a, abd107b, zeraatgari107c}.

Recurrent Neural Networks (RNNs) are neural networks with recurrent connections, defined by an architecture that specifies the number of layers, hidden units, and the recurrence structure for sequential data~\cite{bengio108}. The hidden state \(h_t\) at time step \(t\) is computed as: 
\[
h_t = \sigma(W_h h_{t-1} + W_x x_t + b),
\]
where \(x_t\) is the input at time \(t\), \(W_h\) and \(W_x\) are weight matrices, \(b\) is a bias vector, and \(\sigma\) is an activation function (often \(\tanh\) or ReLU). The learning algorithm, such as backpropagation through time (BPTT), adjusts the weights to capture temporal dependencies for sequence modeling and prediction~\cite{sutskever109, mikolov110}. Uncertainty is often estimated using techniques like dropout. Bias and instability due to vanishing or exploding gradients are mitigated by gradient clipping and batch normalization. Class imbalance is addressed through sequence-level weighting or synthetic time-series generation. Figure~\ref{fig:SL_algorithms}(c) illustrates a typical RNN architecture. RNNs are widely used in astronomy for time-series tasks, including supernova and variable star classification~\cite{charnock249, naul76}, gravitational wave denoising~\cite{shen111a}, periodic variable detection~\cite{tsang216}, and real-time transient discovery~\cite{moller255, muthukrishna252}.

Convolutional Neural Networks (CNNs) are DL models designed for grid-like data such as images and time series~\cite{li123, o124, yamashita125}. A CNN model consists of learnable convolutional filters \(W\) applied to local regions \(X\) of the input, extracting spatially hierarchical features:
\[
(W * X)_{i,j} = \sum_{m,n} W_{m,n} X_{i+m,j+n},
\]
Pooling layers reduce spatial dimensions, improving robustness to translations and distortions, while fully connected layers aggregate extracted features for final predictions. The corresponding learning algorithm adjusts the filter weights and biases via backpropagation and gradient descent.  
Uncertainty can be estimated using dropout or test-time augmentation (TTA). In TTA, multiple augmented versions of the input (e.g., rotations, flips, or shifts) are passed through the trained CNN at inference time, and the predictions are aggregated, for example via averaging or voting. The variability of predictions across these augmentations provides an empirical measure of uncertainty, while the aggregation improves robustness by reducing sensitivity to input transformations and noise.  
Bias is further reduced through data augmentation during training, which improves model invariance. Class imbalance is addressed using loss functions such as focal loss, which downweight well-classified examples. Figure~\ref{fig:SL_algorithms}(g) illustrates a typical CNN architecture. CNNs excel in astronomy tasks such as galaxy morphology classification~\cite{dieleman125a, aniyan125b}, variable star classification~\cite{hon125c}, supernova identification~\cite{brunel250, pasquet254}, photometric redshift estimation~\cite{hoyle125d}, cosmological parameter inference~\cite{ntampaka125e}, and strong gravitational lensing detection~\cite{lanusse125f}.

Long Short-Term Memory (LSTM) networks are a specialized type of RNN, defined by an architecture that includes memory cells and gating mechanisms (input, forget, and output gates) to overcome vanishing and exploding gradient problems~\cite{hochreiter112, le113, gers114}. The cell state \(c_t\) is updated as:
\[
c_t = f_t \odot c_{t-1} + i_t \odot \tanh(W_c [h_{t-1}, x_t] + b_c),
\]
where \(\odot\) denotes element-wise multiplication. The learning algorithm, typically BPTT, adjusts the weights of the gates and hidden units to capture long-term dependencies in sequential data. Figure~\ref{fig:SL_algorithms}(d) illustrates a typical LSTM architecture. LSTMs are used for anomaly detection in LCs~\cite{zhang114a}, classification of solar radio spectra~\cite{xu114b}, detection of core-collapse supernovae~\cite{iess114c}, and solar flare prediction~\cite{sun114d}.

Support Vector Machines (SVMs) are SL models that define a hyperplane separating classes in an \(N\)-dimensional space. The model is characterized by the weight vector \(w\), bias \(b\), and, for non-separable data, slack variables \(\xi_i\). The learning algorithm then identifies the optimal hyperplane by solving the optimization problem: 
\[
\min_{w,b,\xi} \frac{1}{2} \|w\|^2 + C \sum_{i=1}^N \xi_i,
\]
\[
\text{subject to} \quad y_i (w^T x_i + b) \geq 1 - \xi_i, \quad \xi_i \geq 0, \quad i = 1, \dots, N,
\]
where \(C\) is a regularization parameter controlling the trade-off between margin size and classification error. Nonlinear decision boundaries are handled via kernel functions such as the radial basis function (RBF) kernel. Uncertainty can be estimated using Platt scaling~\cite{platt1999probabilistic}, and class imbalance is addressed by adjusting class weights or using synthetic oversampling methods. Figure~\ref{fig:SL_algorithms}(e) illustrates a typical SVM separating hyperplane. SVMs are widely applied in astronomy for star-galaxy separation~\cite{kovacs115}, identification of galaxies~\cite{krakowski116}, classification of strong gravitational lenses~\cite{hartley117}, identification of pre-main-sequence stars~\cite{ksoll119}, and variable stars~\cite{pashchenko120}.

\(k\)-Nearest Neighbors (KNN) are non-parametric, instance-based models that define a prediction as a function of the \(k\) closest training samples according to a distance metric, typically Euclidean distance~\cite{mucherino121, keller122}. For a query point \(x\), the model output is given by:
\begin{align*} 
\hat{y}(x) &= \text{majority vote} \left( \{ y_i \mid x_i \in \mathcal{N}_k(x) \} \right), \\ 
\mathcal{N}_k(x) &= \left\{ \text{the } k \text{ nearest neighbors of } x \right\}, 
\end{align*}
The learning algorithm then determines the neighbors at prediction time using the chosen distance metric. As lazy learners, KNNs require no explicit training but can be computationally expensive for large datasets.  
Uncertainty in KNN predictions can be quantified using distance-weighted voting, where closer neighbors have higher influence on the predicted class, or by computing confidence scores based on the relative distances of neighbors from the query point.   
Feature scaling (normalization or standardization) is important to reduce bias in distance calculations. Class imbalance is addressed by adapting \(k\) or weighting votes to emphasize minority classes. Figure~\ref{fig:SL_algorithms}(f) illustrates a typical KNN structure. KNN is widely applied in astronomy for classifying celestial objects~\cite{li122a}, photometric redshift estimation~\cite{zhang122b}, star–galaxy separation~\cite{machado122c}, and solar flare prediction~\cite{li122d}.

\subsection{Unsupervised Learning}\label{subsec:Unsupervised}

\begin{figure*}[htbp]
    \centering
    \includegraphics[width=0.9\textwidth]{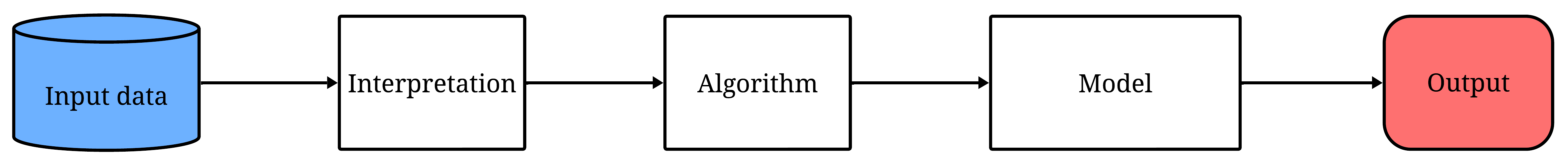}
    \caption{Basic process of UL. Cylinder represents database, rectangles indicate data processing steps and rounded rectangle shows the final output.}
    \label{fig:UL}
\end{figure*}

Unsupervised learning (UL) is a ML approach that trains on unlabeled data, identifying underlying patterns and structures without predefined labels. Unlike supervised methods, which rely on labeled examples, unsupervised algorithms autonomously detect relationships, group similar data points, and uncover anomalies without external guidance. The basic flow of UL is shown in Figure~\ref{fig:UL}.  

UL encompasses a broad range of statistical techniques for data exploration, including clustering, dimensionality reduction, visualization, and anomaly detection. These methods are especially valuable in scientific research, as they enable the discovery of hidden patterns and the extraction of new insights from large datasets.

\begin{figure*}[htbp]
    \centering
   \includegraphics[width=0.9\textwidth]{Figures/UL_algorithms.pdf}
    \caption{Overview of UL algorithms. (a) K-means: Partitions data into \(k\) clusters by minimizing the variance within each cluster. (b) HC: Hierarchical clustering builds a hierarchy of clusters. (c) PCA: Reduces dimensionality by projecting data onto orthogonal axes that maximize variance, with the first principal component shown. (d) t-SNE: Embeds high-dimensional data into a low-dimensional space while preserving local structures. (e) AE: An autoencoder consisting of an encoder and decoder, used to learn a low-dimensional latent representation of the input data. (f) SOM: A self-organizing map that projects high-dimensional data onto a low-dimensional grid while preserving topological properties.}
    \label{fig:UL_algorithms}
\end{figure*}

K-means is a widely used clustering model valued for its simplicity and efficiency. The model partitions data into $K$ clusters, each represented by a centroid $\mu_k$, and aims to minimize the within-cluster variance:
\[
\min_{\{C_k\}} \sum_{k=1}^K \sum_{x_i \in C_k} \|x_i - \mu_k\|^2,
\]
The corresponding K-means learning algorithm initializes $K$ cluster centroids randomly, then iteratively assigns points to the nearest centroid and updates centroids until convergence~\cite{macqueen126,hartigan127, sinaga128}. Uncertainty is addressed through soft clustering variants, while bias is mitigated via careful centroid initialization. K-means is sensitive to class imbalance but can be adapted using density-based weighting. Figure~\ref{fig:UL_algorithms}(a) illustrates how K-means works, showing the iterative process of centroid updates and cluster assignments. K-means is used for classifying stellar spectra~\cite{almeida128a}, distinguishing stellar clusters based on chemical abundances~\cite{garcia128b}, determining membership probabilities in open clusters~\cite{el128c}, and classifying eclipsing binary LCs~\cite{modak219}.

Hierarchical clustering (HC) is a popular clustering model that builds a hierarchy of clusters~\cite{ward129}. There are two main types of HC: Agglomerative Hierarchical Clustering (AHC) and Divisive Hierarchical Clustering (DHC). AHC, also known as the "bottom-up" approach, models each data point as an individual cluster and defines a merging criterion based on a chosen distance metric. DHC, or the "top-down" approach, models all data points initially as a single cluster and defines a splitting criterion to recursively divide clusters until each point becomes its own cluster~\cite{murtagh130, johnson131}.  
The HC learning algorithm builds the cluster hierarchy using linkage criteria:
\[
d(C_i,C_j) = \begin{cases}
\min_{x \in C_i,\, y \in C_j} d(x,y) & \text{(single linkage)}, \\
\max_{x \in C_i,\, y \in C_j} d(x,y) & \text{(complete linkage)},
\end{cases}
\]
Cophenetic correlation quantifies uncertainty, and linkage choice affects bias. Figure~\ref{fig:UL_algorithms}(b) illustrates AHC, showing the process of merging clusters in a "bottom-up" manner. Applications of HC include identifying galaxy cluster substructures~\cite{yu131a}, detecting superclusters of galaxies~\cite{santiago131b}, exploring the hierarchical structure of open clusters~\cite{yu131c}, and clustering astronomical spectra~\cite{tian131d}.

Principal Component Analysis (PCA) is a dimensionality reduction model that transforms high-dimensional data into a lower-dimensional representation while retaining as much relevant information as possible~\cite{wold132}. The model identifies orthogonal directions, called principal components, which capture the maximum variance in the dataset. Projecting data onto these components reduces the number of features, mitigates the curse of dimensionality, enhances computational efficiency, and improves interpretability~\cite{abdi133, jolliffe134}.  

The PCA learning algorithm computes the eigenvectors of the covariance matrix and projects data onto the top-$k$ eigenvectors:
\[
X' = X V_k, \quad \Sigma = X^T X,
\]
where $V_k$ contains the top-$k$ eigenvectors. Reconstruction error estimates uncertainty, and standardization prevents feature scale bias. Figure~\ref{fig:UL_algorithms}(c) demonstrates the PCA process. PCA is applied in astronomy for searching galaxy-scale strong lenses~\cite{joseph134a}, exploring abundances in different environments~\cite{ting134b}, and analyzing LCs of variable stars~\cite{deb134c}.

t-Distributed Stochastic Neighbor Embedding (t-SNE) is a dimensionality reduction model designed to represent high-dimensional data in two or three dimensions while preserving local similarity relationships~\cite{van135}. The model maps similar objects close together and dissimilar objects farther apart, capturing non-linear relationships and revealing clusters and intricate patterns in complex datasets~\cite{van136, kobak137}.  
The t-SNE learning algorithm minimizes the Kullback-Leibler divergence between probability distributions in the high- and low-dimensional spaces:
\[
KL(P \| Q) = \sum_i P(i) \log \frac{P(i)}{Q(i)},
\]
where the perplexity parameter balances local and global structure, affecting uncertainty. Careful tuning is required to avoid initialization bias. Figure~\ref{fig:UL_algorithms}(d) illustrates the t-SNE process. t-SNE is applied in astronomy to identify SED outliers~\cite{quispe137a}, reduce spectral information~\cite{traven137b}, quantify similarity between planetary systems~\cite{alibert137c}, and separate Type-2 AGN from HII galaxies~\cite{zhang137d}.

Autoencoder (AE) is an ANN model designed to learn efficient low-dimensional representations of input data, commonly used for tasks such as compression, dimensionality reduction, and visualization~\cite{yang138, gianniotis139, gianniotis98}. The model consists of two main components: an encoder, which compresses the input data into a lower-dimensional latent representation, and a decoder, which reconstructs the original data from this compressed form.  
The AE learning algorithm optimizes network weights by minimizing the reconstruction error, typically measured as the squared difference between the input and reconstructed output:
\[
\mathcal{L} = \| x - g_\theta(f_\phi(x)) \|^2,
\]
where $f_\phi$ encodes and $g_\theta$ decodes. Dropout layers estimate uncertainty, while regularization prevents overfitting. Figure~\ref{fig:UL_algorithms}(e) shows the structure of an AE. AEs are applied in astronomy to estimate gravitational wave parameters~\cite{gabbard139a}, identify strong gravitational lenses~\cite{cheng139b}, estimate atmospheric parameters~\cite{yang138}, and detect anomalies in cosmic ray variations~\cite{mandrikova139d}.

Self-organizing map (SOM), also known as a Kohonen map~\cite{kohonen140}, is an unsupervised ANN model that creates a low-dimensional, typically two-dimensional, representation of high-dimensional input data. The model represents the output neurons with weight vectors that act as prototypes or templates, capturing the essential features of the input dataset. This structure allows SOMs to effectively map and visualize high-dimensional data in a more interpretable, lower-dimensional form while preserving the input topology.
The SOM learning algorithm organizes the map by minimizing the quantization error:
\[
Q = \sum_i \| x_i - w_{c(i)} \|^2.
\]
where $w_{c(i)}$ is the winning neuron. Quantization error measures uncertainty, while neighborhood preservation reduces topological bias. Figure~\ref{fig:UL_algorithms}(f) illustrates the SOM process. SOMs are applied in astronomy to estimate galaxy photometric redshift probability density functions (P

\subsection{Semi-supervised Learning}\label{subsec:Semi-supervised}

Semi-supervised learning (SSL) is a ML approach that integrates elements of both SL and UL. It leverages a small amount of labeled data alongside a large pool of unlabeled data, making it particularly useful when labeled data is scarce but unlabeled data is abundant~\cite{chapelle145}.  

\begin{figure*}[htbp]
    \centering
   \includegraphics[width=0.9\textwidth]{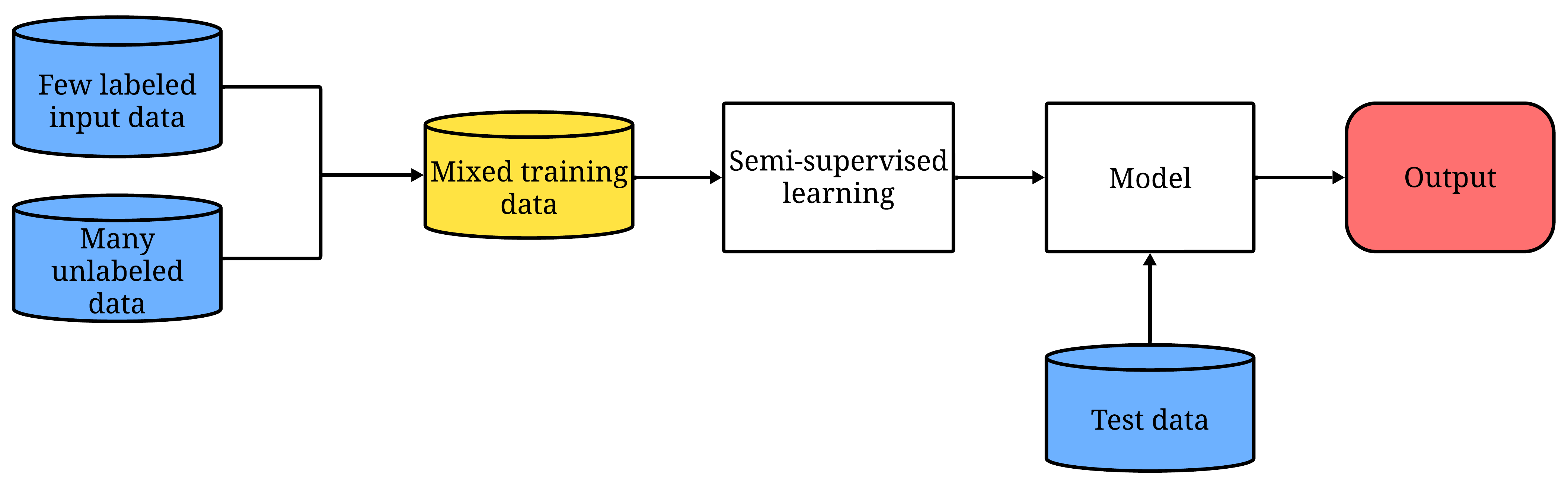}
    \caption{Basic process of SSL. Cylinders represent databases, rectangles indicate data processing steps, and the rounded rectangle shows the final output.}
    \label{fig:SSL}
\end{figure*}

The SSL process begins with dataset collection, comprising both labeled and unlabeled data, followed by cleaning and preprocessing to ensure consistency. The model is initially trained on the labeled data to establish a foundational understanding of the task. It then refines its performance by incorporating information from the unlabeled data, improving overall accuracy. The general workflow of SSL is illustrated in Figure~\ref{fig:SSL}. 

SSL can be categorized into \textit{inductive} and \textit{transductive} learning methods~\cite{van146}. Inductive learning aims to develop a generalized model capable of making predictions on unseen data. It utilizes both labeled and unlabeled data during training to enhance generalization. In contrast, transductive learning focuses on predicting labels solely for the specific unlabeled data available during training, without aiming for broader generalization. Transductive methods often exploit the inherent structure of the unlabeled data, such as relationships between data points, to improve prediction accuracy.  

SSL offers several advantages, including more efficient data utilization by leveraging both labeled and unlabeled data, reducing the cost associated with manual labeling. Additionally, it enhances model performance by capturing structural patterns in unlabeled data. By bridging the gap between SL, which relies on labeled data, and UL, which identifies patterns without predefined labels, SSL provides a robust framework for learning in scenarios with limited labeled datasets.

SSL is used in various astronomical tasks, such as classifying photometric supernovae~\cite{richards62}, classifying radio galaxies~\cite{slijepcevic146a}, detecting extrasolar planets~\cite{sulis146b}, classifying and clustering variable stars~\cite{pantoja231}, identifying pulsar candidates~\cite{balakrishnan146c}, classifying galaxy morphologies and detecting anomalies across surveys~\cite{ciprijanovic146d}, verifying molecular clumps~\cite{luo146e}, distinguishing real and bogus transients~\cite{liu146f}, and detecting solar filaments in H$\alpha$ observations~\cite{diercke146g}.

\subsection{Uncertainty and Bias}
In ML models, uncertainty arises from two primary sources. Epistemic uncertainty (model uncertainty) reflects the lack of knowledge due to limited or unrepresentative training data and can be reduced by collecting additional observations or using ensemble or Bayesian methods. For example, a CNN trained on a limited set of supernova LCs may be uncertain when classifying rare or novel transients. Aleatoric uncertainty (data uncertainty) originates from inherent noise in measurements, such as photometric errors or instrumental noise, and cannot be reduced by more data. Bias can also occur systematically due to model assumptions or dataset composition; for instance, overrepresentation of bright galaxies may lead to misclassification of faint objects. Characterizing both uncertainty and bias is crucial in astronomy, as it informs the reliability of classifications, the confidence in parameter estimates, and the prioritization of follow-up observations. Techniques such as Monte Carlo dropout, Bayesian neural networks, and careful data augmentation are employed to quantify and mitigate these effects across the models.

\subsection{Evaluation Metrics and Validation}\label{sec:Evaluation}
\subsubsection{Evaluation Metrics}

Evaluation metrics are essential tools in ML that quantitatively assess a model's performance. They play a crucial role in optimizing hyper-parameters, evaluating model effectiveness, selecting the most relevant features, and comparing different ML algorithms. These metrics are computed during both the validation and testing phases, where the trained model is applied to previously unseen data, and its predictions are compared against actual target values to measure accuracy and reliability.

In regression problems, where the target variable is continuous, model performance is commonly evaluated using metrics such as the Mean Absolute Error (MAE) and the Mean Squared Error (MSE). The MAE is defined as:
\[
\text{MAE} = \frac{1}{n} \sum_{i=1}^{n} |y_i - \hat{y}_i|,
\]
where \( n \) is the number of instances in the validation or test set, \( y_i \) is the true value, and \( \hat{y}_i \) is the model's prediction. The MSE is given by:
\[
\text{MSE} = \frac{1}{n} \sum_{i=1}^{n} (y_i - \hat{y}_i)^2.
\]

Moreover, additional evaluation metrics are also found in the literature, such as the Normalized Median Absolute Deviation, the Continuous Ranked Probability Score, and the Probability Integral Transform (see e.g.,~\cite{d147a, d83}).

In classification tasks, where the target variable is discrete, commonly used evaluation metrics include Classification Accuracy, the Confusion Matrix, and the Area Under the ROC Curve (AUC).

Classification accuracy is defined as the ratio of correctly predicted labels to the total number of predictions:
\[
\text{Accuracy} = \frac{\text{Number of correct predictions}}{\text{Total predictions}},
\]
with values ranging between 0 and 1. It is most appropriate when classes are balanced and misclassifications are equally important.

\begin{figure}[htbp]
    \centering
    \includegraphics[width=0.95\linewidth]{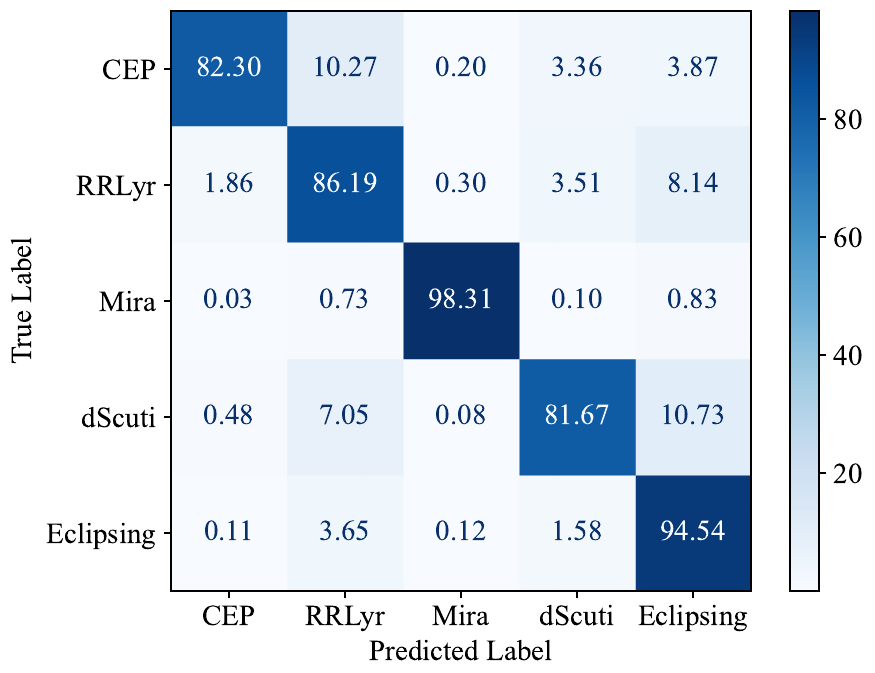}
    \caption{Example of a confusion matrix taken from~\cite{akhmetali244}, who trained a CNN model to distinguish between 5 classes of variable stars. The confusion matrix displays the percentage distribution of true class labels against the predicted class labels. In the best-case scenario, the confusion matrix is purely diagonal, with all off-diagonal elements equal to zero, indicating that all instances were correctly classified.}
    \label{fig:CM}
\end{figure}

The confusion matrix provides a more detailed view of performance, especially in multi-class settings. Figure~\ref{fig:CM} shows an example of a confusion matrix, taken from~\cite{akhmetali244}, who trained a CNN model to distinguish between 5 classes of variable stars. Each row corresponds to the true class, and each column to the predicted class. Ideally, the matrix is diagonal, indicating perfect classification. Normalizing the matrix removes dependence on class size, allowing for easier comparison across categories.

The Receiver Operating Characteristic (ROC) curve is a standard diagnostic tool for binary classification. It plots the True Positive Rate (TPR) against the False Positive Rate (FPR):
\[ \text{TPR} = \frac{\text{True Positives}}{\text{Total Actual Positives}},\] \[ \text{FPR} = \frac{\text{False Positives}}{\text{Total Actual Negatives}}. \]

\begin{figure}[htbp]
    \centering
    \includegraphics[width=0.95\linewidth]{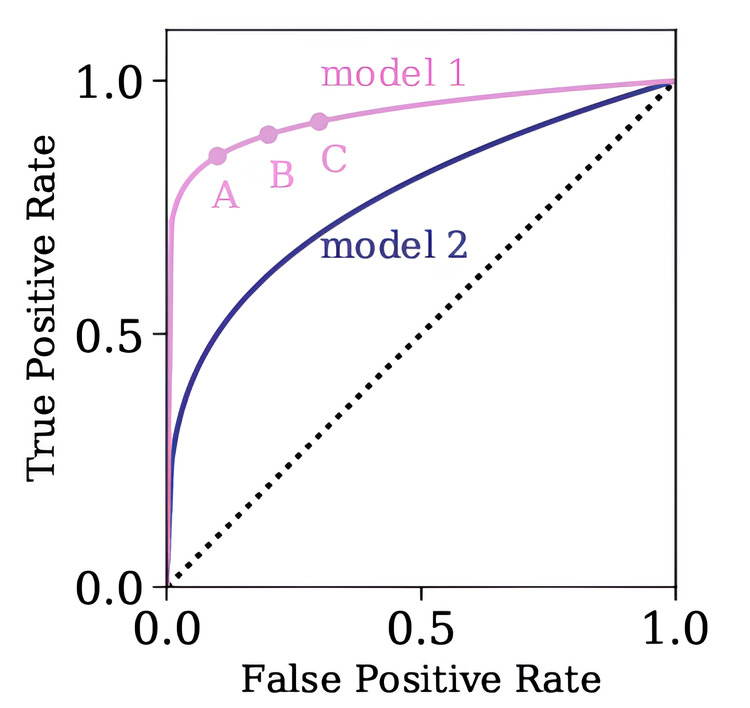}
    \caption{An illustration of a ROC curve taken from~\cite{baron10}, which plots the TPR against the FPR. Ideally, a model should achieve a TPR of 1 with a FPR of 0. The black dashed line represents the performance of a random classifier. By adjusting the model's hyperparameters, different points on the ROC curve are generated using validation set results. The pink curve corresponds to model 1, with points A, B, and C indicating specific hyperparameter settings. The purple curve shows the performance of model 2. The area under the ROC curve (AUC) provides a measure for model comparison - model 1 performs better in this case.}
    \label{fig:ROC}
\end{figure}

By varying the classification threshold, the ROC curve visualizes model performance across different operating points. The Area Under the Curve (AUC) provides a summary measure of overall performance, with higher AUC values indicating better discrimination. ROC curves are valuable for comparing models and for guiding the selection of classification thresholds that balance sensitivity and specificity, as illustrated in Figure~\ref{fig:ROC}.

In multiclass settings, ROC analysis requires adaptation because TPR and FPR are inherently defined for binary outcomes. One approach is the one-vs-rest (OvR) strategy, in which each class is treated as the positive class while all other classes are treated as negative, producing a separate ROC curve for each class. Alternatively, macro-averaging computes the AUC for each class and averages them equally, giving each class the same weight, while micro-averaging aggregates true positives, false positives, and false negatives across all classes before calculating TPR and FPR, effectively weighting classes by prevalence. Interpreting ROC results in multiclass problems can be challenging, particularly when dealing with imbalanced datasets or more than three classes, and visualizing multiple ROC curves simultaneously may be difficult.

As an alternative to ROC, Precision-Recall (PR) curves are often more informative for imbalanced datasets because they emphasize performance on the positive class. PR curves can also guide threshold selection by highlighting points that optimize the trade-off between precision and recall.

By analyzing ROC or PR curves, it is possible to choose classification thresholds that optimize specific metrics such as the F1 score, balanced accuracy, or sensitivity, rather than relying on a default threshold of 0.5. This is especially important in astronomy, where rare events, such as supernovae or transient detections, require prioritizing sensitivity to avoid missing important signals.

\subsubsection{Validation Methods}\label{sec:Validation}

To reliably assess the generalization ability of ML models, it is crucial to evaluate them on data that was not used during training. This is achieved through various validation methods, which help prevent overfitting and provide unbiased estimates of model performance.

Holdout Validation (HV)~\cite{kim2009estimating} is the most common approach for evaluating ML models. In this approach, the dataset is split into training, validation, and test sets. A typical split uses 70\% for training, 15\% for validation, and 15\% for testing, though these ratios may vary depending on dataset size and variability. For small datasets, a larger training proportion is preferred to ensure sufficient learning. As datasets grow, smaller validation sets can still capture data diversity effectively. The training set is used to fit model parameters, while the validation set is used for hyperparameter tuning—a process that, although computationally intensive, can be automated and parallelized. The final model is evaluated on the unseen test set. HV is used in various astronomical tasks, such as classification of spectral data~\cite{solorio2023random}, estimation of galaxy redshifts~\cite{ball2008robust}, classification of sterllar variability~\cite{audenaert2021tess}, and predicting stellar rotations~\cite{gomes2024predicting}.

Cross-validation (CV)~\cite{stone1974cross} is a resampling technique used to obtain an unbiased estimate of a model’s performance. Compared to simple holdout validation, it generally offers a more reliable assessment of a model’s generalization error, particularly when working with small datasets. Several cross-validation strategies commonly used for model evaluation are outlined below. CV is used in astronomy for estimation of photometric redshifts of quasars~\cite{ball2007robust}, Star--Galaxy--QSO classification~\cite{bai2018machine}, calibration of ages for the APOGEE DR17 catalogue~\cite{boulet2024catalogue}, and classification of variabile sources~\cite{richards2012construction}.

K-Fold Cross-Validation (KFCV)~\cite{bengio2004no} partitions the dataset randomly into \(k\) disjoint subsets, or folds. In each iteration, one fold is held out as the validation set, while the remaining \(k-1\) folds are used for training. This process repeats \(k\) times, ensuring each fold serves exactly once as the validation set. The average of the performance metrics across all iterations provides a robust estimate of the model’s validation error. KFCV is used in various works, such as classification of variable stars~\cite{akhmetali244}, star--galaxy separation~\cite{machado2016exploring}, protostellar classification~\cite{miettinen2018protostellar}, and identification and fitting eclipse maps of exoplanets~\cite{hammond2024identifying}.

Stratified K-Fold Cross-Validation (SKFCV)~\cite{prusty2022skcv} is particularly useful in classification tasks with imbalanced class distributions. Unlike standard KFCV, stratified KFCV preserves the original class proportions within each fold. This ensures that minority classes are adequately represented, yielding more stable and representative performance estimates. SKFCV is used for prediction of habitability of exoplanets~\cite{raminaidu2023building}, classification of massive stars in nearby galaxies~\cite{maravelias2022machine}, spectral classification of OB stars~\cite{kyritsis2022new}, and detection of quasi-periodic oscillations in X-ray binaries~\cite{kiker2023qpoml}.

Leave-One-Out Cross-Validation (LOOCV)~\cite{wong2015performance} is a special case of KFCV where \(k = n\), the total number of data points. Each iteration uses a single sample for validation and the remaining \(n-1\) samples for training. LOOCV provides nearly unbiased performance estimates, making it well-suited for very small or imbalanced datasets. However, it is computationally intensive and often impractical for large datasets or models with long training times. SKFCV is for exoplanet atmospheric analysis~\cite{welbanks2023application}, selection of optically variable active galactic nuclei~\cite{de2025selection}, analysis of stellar LCs~\cite{pan243}, and exoplanet detection~\cite{pimentel2024feature}.

Nested Cross-Validation (NCV)~\cite{varma2006bias} addresses the challenge of reliable hyperparameter tuning and unbiased performance estimation. The data is split into outer and inner loops of cross-validation. The inner loop is used for model selection (e.g., tuning hyperparameters), while the outer loop assesses the generalization error. This method helps avoid optimistic bias that can occur when hyperparameters are tuned and evaluated on the same data. NCV is used for star/galaxy classification~\cite{chao2020research}, cosmological inference from non-linear scales~\cite{yuan2024robust}, classification of variable stars~\cite{long2012optimizing}, and prediction of galaxy metallicity from three-colour images~\cite{wu2019using}.

Validation methods not only influence model selection but also help assess uncertainty, bias, and variance. For instance, large variations in performance across folds may indicate that the model is sensitive to the training data and prone to overfitting. Similarly, if the model consistently underperforms across validation sets, it may be underfitting or affected by biased assumptions.

Despite the common use of performance metrics and validation schemes, explicit statistical comparisons between models are still uncommon in astronomy. This is problematic because reported differences in metrics (e.g., accuracy or AUC) may not be statistically significant, especially with noisy or imbalanced data. Statistical significance tests help assess whether performance improvements are robust rather than due to random fluctuations~\cite{demvsar2006statistical}. For instance, McNemar’s test~\cite{mcnemar1947note} evaluates paired classification outcomes, while paired $t$-tests or Wilcoxon signed-rank tests~\cite{higgins2004introduction} can compare cross-validation scores. Bootstrap and permutation methods can also provide confidence intervals for performance differences~\cite{good2005permutation}. Incorporating such tests offers a more rigorous foundation for model selection and helps prevent overstatement of results.

\section{Applications of Machine Learning in Light Curve Analysis}\label{sec:Application}

The increasing data volume from large-scale astronomical surveys such as \textit{Kepler}, \textit{TESS}, and the upcoming \textit{LSST} presents both unprecedented opportunities and significant challenges in processing and interpretation. Traditional methods struggle with the scale, complexity, and noise in LC data, making automation essential. ML has emerged as a transformative approach, enabling efficient classification, detection, and characterization of astronomical objects with remarkable accuracy.

ML techniques are particularly effective at handling the high-dimensionality and noise inherent in LC data. By leveraging SL, UL, and SSL algorithms, ML facilitates pattern recognition, object classification, and anomaly detection with greater efficiency and precision than traditional methods. This section explores three pivotal applications of ML in LC analysis: (1) exoplanet detection, (2) variable star analysis, and (3) supernova classification.

\subsection{Transiting Exoplanet Detection}\label{sec:Exoplanet}

\begin{figure}[htbp]
    \centering
    \includegraphics[width=0.95\linewidth]{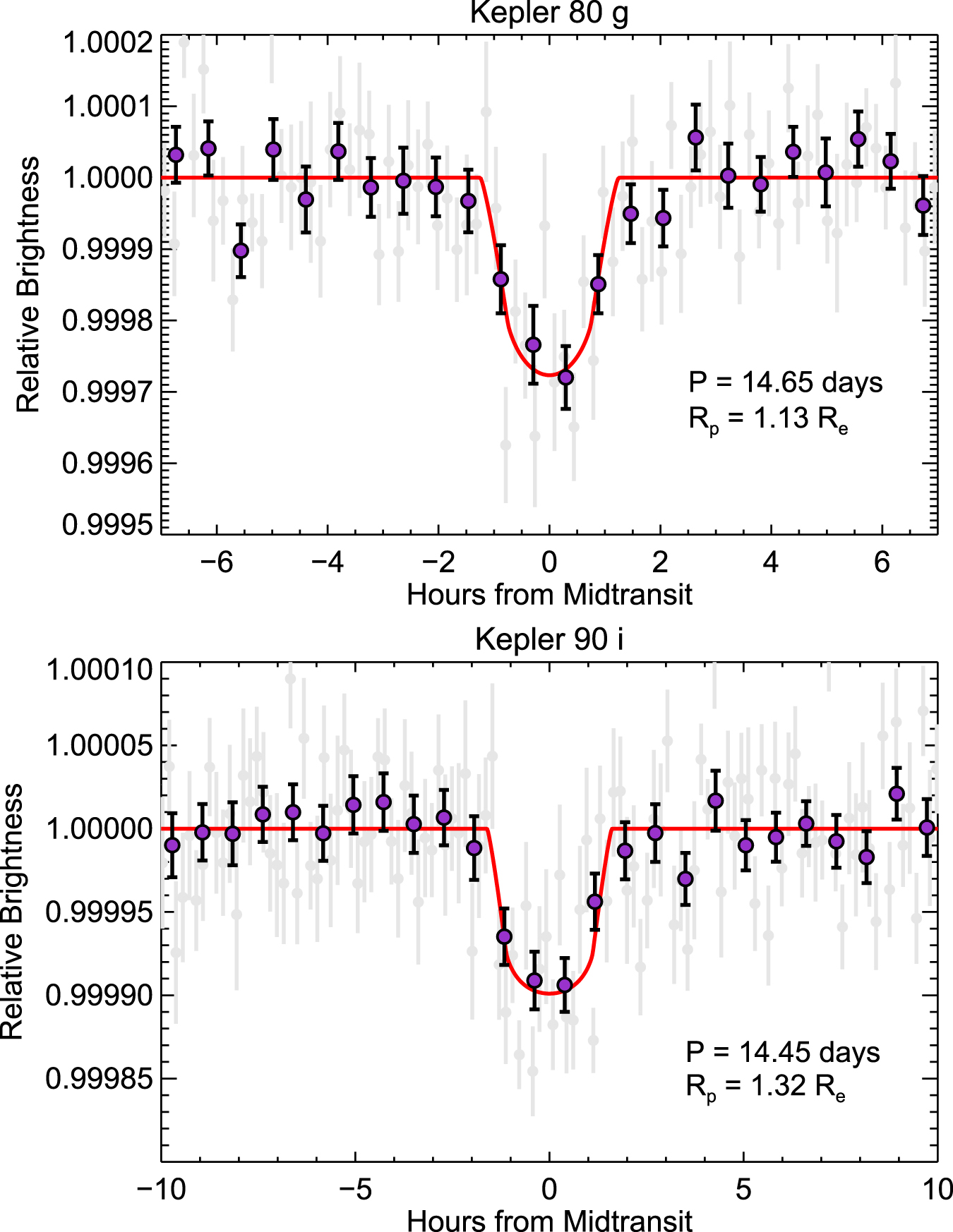}
    \caption{Example of transiting exoplanets. The figure is sourced from~\cite{shallue149}.}
    \label{fig:Transit}
\end{figure}

Transit photometry has become a fundamental technique for exoplanet discovery, detecting exoplanets by identifying periodic dips in a star’s brightness and serving as a cornerstone of modern planet-hunting surveys. Figure~\ref{fig:Transit} illustrates a typical LC with a transit event, showcasing the subtle flux dip that ML algorithms are trained to recognize. 

Space-based missions including \textit{Kepler} (2009-2018) and \textit{TESS} (2018-present) have revolutionized this field by providing high-precision photometric data across large sky areas. The \textit{Kepler} mission identified thousands of planetary candidates through continuous monitoring of a single field, while \textit{TESS} has expanded this catalog using its all-sky survey strategy focused on brighter stars. The analysis of these datasets presents significant computational challenges due to their volume, noise characteristics, and the presence of astrophysical false positives, necessitating advanced analytical approaches. Table~\ref{table2} presents a summary of ML methods applied to different exoplanet detection datasets.

The \textit{Kepler} mission data have served as a testbed for developing ML techniques in transit detection. Initial work established the effectiveness of CNNs through the \texttt{AstroNet} architecture~\cite{shallue149}, which achieved classification performance comparable to human experts. Subsequent developments introduced modifications such as \texttt{ExoNet}~\cite{ansdell150}, incorporating additional diagnostic information to reduce false positives, and \texttt{AstroNet-K2}~\cite{dattilo151}, adapted for the modified observing strategy of the K2 mission. Alternative approaches including 2D-CNN architectures~\cite{chintarungruangchai152} and ensemble methods~\cite{priyadarshini153} demonstrated improved sensitivity to low signal-to-noise transits.

Analysis of \textit{TESS} data has built upon these foundations while addressing the mission's distinct characteristics. Modified versions of the original \texttt{AstroNet} framework have been applied to \textit{TESS} observations~\cite{yu170}, with subsequent refinements such as \texttt{Astronet-Triage-v2}~\cite{tey174} improving classification accuracy. The \texttt{Nigraha} pipeline~\cite{rao172} represents a comprehensive implementation integrating multiple analysis stages, while systems like \texttt{SHERLOCK}~\cite{devora177} provide end-to-end processing capabilities. The \texttt{ExoMiner}~\cite{valizadegan178} and its enhanced version \texttt{ExoMiner++} ~\cite{valizadegan179} have demonstrated the potential of DL to replicate and augment expert vetting processes.

ML applications have been successfully adapted to various astronomical surveys and simulated datasets beyond the primary space-based missions. In simulated LC analysis, CNNs were applied to detect transits of habitable planets in high-cadence data~\cite{zucker79}, while alternative approaches included 1D CNNs for processing non-phase-folded LCs~\cite{iglesias181}. GPU-accelerated phase-folding algorithms were developed specifically for detecting ultrashort-period exoplanets~\cite{wang182}.

For ground-based surveys, different ML approaches were implemented: RF and SOM techniques were combined in the NGTS survey for candidate vetting~\cite{armstrong183}, and CNNs were employed for automated candidate screening~\cite{chaushev184}. The QES project utilized DBSCAN-based algorithms for effective noise rejection in transit data~\cite{mislis185}. Similarly, the WASP survey integrated RF and CNN methods for comprehensive transit signal analysis~\cite{schanche186}.

CNNs were also implemented for the BRITE mission's photometric data analysis~\cite{yeh187}. For infrared observations, LSTM networks were applied to Spitzer data for improved detrending of LCs~\cite{morvan188}. In \textit{Gaia} photometry, XGBoost-assisted methods were developed for transit searches, leading to confirmed exoplanet discoveries~\cite{panahi189}.

These diverse applications demonstrate the versatility of ML techniques across various observational platforms, data types, and specific scientific requirements. The systematic development of analysis methods, reflects the ongoing evolution of techniques to address the challenges posed by current and future transit surveys. These methodological advances continue to enhance the detection and characterization of exoplanetary systems across diverse observational datasets.

\begin{longtable*}{|m{3cm}|m{4cm}|m{10cm}|}
    \caption{Summary of ML methods applied to exoplanet detection datasets.}
    \label{table2} \\
    \hline
    \textbf{Data type} & \textbf{Method} & \textbf{Description} \\ \hline 
    \endfirsthead
    
    \multicolumn{3}{c}{{\tablename\ \thetable{} - Continued}} \\
    \hline
    \textbf{Data type} & \textbf{Method} & \textbf{Description} \\ \hline 
    \endhead
    
    \hline
    \endfoot
    
    \hline
    \endlastfoot
    
    \multirow{11}{*}{Kepler} 
    & \multirow{11}{*}{CNN}
    &~\cite{pearson148} proposed a CNN-based method for exoplanet detection,   outperforming least-squares techniques without requiring model fitting. \\ \cline{3-3}
    &&~\cite{shallue149} developed \texttt{AstroNet}, a deep CNN designed for   exoplanet classification. \\ \cline{3-3}
    &&~\cite{ansdell150} proposed \texttt{ExoNet}, extending \texttt{AstroNet} by incorporating   domain knowledge, centroid time-series data, and stellar parameters. \\ \cline{3-3}
    &&~\cite{dattilo151} proposed \texttt{AstroNet-K2}, an extension of \texttt{AstroNet} adapted for Kepler's K2 data. \\ \cline{3-3}
    &&~\cite{chintarungruangchai152} proposed a 2D-CNN model with   phase-folding for transit detection, demonstrating improved accuracy at low S/N. \\ \cline{3-3}
    &&~\cite{priyadarshini153} proposed an Ensemble-CNN model for   exoplanet detection, comparing its performance with various ML algorithms. \\ \cline{3-3}
    &&~\cite{bugueno154} proposed a CNN-based exoplanet detection method using   MTF to transform unevenly sampled LCs into fixed-size images. \\ \cline{3-3}
    &&~\cite{cuellar155} proposed a CNN-based transit detection model trained   on mixed real and synthetic data. \\ \cline{2-3}
    
    & \multirow{6}{*}{RF}
    &~\cite{jenkins156} proposed a RF-based approach to automate transit signal   classification, generating a preliminary list of planetary candidates. \\ \cline{3-3}  
    &&~\cite{mccauliff157} expanded RF-based exoplanet classification by   transforming transit-like detections into numerical attributes. \\ \cline{3-3}
    &&~\cite{sturrock158} developed an RF-based exoplanet classification model   and deployed it as a publicly accessible API in the cloud. \\ \cline{3-3} 
    &&~\cite{caceres159} developed the ARPS method combining ARIMA modeling,   transit comb filtering, and RF classification to identify exoplanet candidates. \\ \cline{3-3}
    &&~\cite{jin160} optimized SL with feature selection and tuning, with RF   performing best, and used clustering to identify potentially habitable exoplanets. \\ \cline{3-3}
    &&~\cite{hesar161} evaluated six classification algorithms for exoplanet detection,   identifying RF and SVM as the top performers based on accuracy and F1 score. \\ \cline{2-3}
    
    & \multirow{1}{*}{KNN}
    &~\cite{thompson162} proposed a ML-based metric using dimensionality   reduction and KNN to identify transit-shaped signals. \\ \cline{3-3}
    &&~\cite{bahel163} explored exoplanet detection using ML classification,   applying KNN on SMOTE-balanced data. \\ \cline{2-3}

    & \multirow{1}{*}{Ensemble}
    &~\cite{hesar164} applied ML models to estimate stellar rotation periods,   demonstrating that Voting Ensemble improves accuracy over traditional approaches. \\ \cline{3-3}
    &&~\cite{luz165} evaluated five Ensemble ML algorithms for exoplanet classification. \\ \cline{2-3}
    
    & ANN &~\cite{kipping166} developed an ANN-based model to predict short-period   transits likely to have additional planets. \\ \cline{2-3}
    & SOM &~\cite{armstrong142} developed a SOM-based method for fast exoplanet   candidate classification. \\ \cline{2-3}
    & GPC &~\cite{armstrong167} proposed a GPC-based probabilistic planet validation   method as an alternative to VESPA. \\ \cline{2-3}
    & LightGBM &~\cite{malik168} proposed a ML approach using TSFresh-extracted features and   a gradient boosting classifier for transit detection. \\ \cline{2-3}
    & GAN &~\cite{suresh169} explored GAN-based data augmentation for exoplanet detection,   showing comparable accuracy with synthetic data and improved performance. \\ 
    \hline
    
    \multirow{7}{*}{TESS} 
    & \multirow{7}{*}{CNN} 
    &~\cite{yu170} modified \texttt{AstroNet} for automated triage and vetting of   TESS candidates. \\ \cline{3-3}
    &&~\cite{osborn171} adapted \texttt{ExoNet} for TESS data, training on simulated LCs. \\ \cline{3-3}
    &&~\cite{rao172} developed \texttt{Nigraha}, built upon \texttt{AstroNet}, a pipeline combining   transit detection, supervised ML, and vetting. \\ \cline{3-3}
    &&~\cite{olmschenk173} developed a CNN for efficient exoplanet transit   detection, and identified 181 new exoplanet candidates. \\ \cline{3-3}
    &&~\cite{tey174} developed \texttt{Astronet-Triage-v2}, built upon \texttt{AstroNet},   an improved neural network for exoplanet candidate triage. \\ \cline{3-3}
    &&~\cite{fiscale175} demonstrated that combining transfer learning with   regularization techniques significantly enhances CNN performance. \\ \cline{3-3}
    &&~\cite{liao176} proposed a wavelet-transform-based LC representation   and an improved Inception-v3 CNN. \\ \cline{3-3}
    &&~\cite{devora177} proposed \texttt{SHERLOCK}, an end-to-end pipeline that   enables efficient exoplanet searches. \\ \cline{2-3}
    
    & \multirow{2}{*}{DNN}
    &~\cite{valizadegan178} proposed \texttt{ExoMiner}, a DL classifier that   mimics expert vetting for transit signals. \\ \cline{3-3}
    &&~\cite{valizadegan179} introduced \texttt{ExoMiner++}, an enhanced version of \texttt{ExoMiner},  improving transit signal classification by integrating transfer learning. \\ \cline{2-3}
    
    & Transformer &~\cite{salinas180} proposed a Transformer-based NN for exoplanet detection,   identifying transit signals without phase folding or periodicity assumptions. \\ 
    \hline
    
    \multirow{6}{*}{Simulated data} 
    & \multirow{6}{*}{CNN}
    &~\cite{zucker79} proposed a CNN-based approach to detect transits of   habitable planets in simulated high-cadence LCs. \\ \cline{3-3}
    &&~\cite{iglesias181} developed a 1D CNN for detecting transits in   non-phase-folded LCs. \\ \cline{3-3}
    &&~\cite{wang182} proposed \texttt{GPFC}, a GPU-accelerated phase-folding algorithm   for ultrashort-period exoplanet detection. \\ \hline
    
    \multirow{1}{*}{NGTS}  
    & RF and SOM &~\cite{armstrong183} developed \texttt{autovet}, a ML pipeline, combining   RFs and SOMs to rank planetary candidates with high accuracy. \\ \cline{2-3} 
    & CNN &~\cite{chaushev184} applied a CNN for automated vetting of exoplanet   candidates, reducing manual effort. \\ \hline
    
    QES & DBSCAN &~\cite{mislis185} developed \texttt{TSARDI}, a DBSCAN-based UL algorithm for   noise rejection in transit surveys. \\ \hline
    WASP & RF and CNN &~\cite{schanche186} developed a ML pipeline combining RFs and CNNs   for automated vetting of transit signals. \\ \hline
    BRITE & CNN &~\cite{yeh187} applied CNNs to BRITE LCs for exoplanet transit detection. \\ \hline
    Spitzer Space Telescope & LSTM &~\cite{morvan188} proposed \texttt{TLCD-LSTM}, a probabilistic   LSTM-based detrending method for transit LCs. \\ \hline
    \textit{Gaia} & XGBoost &~\cite{panahi189} developed a ML-assisted transit search in \textit{Gaia} photometry,   leading to the first exoplanet detections by \textit{Gaia}, confirmed as hot Jupiters   via radial velocity measurements. \\ \hline
\end{longtable*}

\subsection{Variable Star Analysis}\label{sec:Variable}

Stellar variability is a fundamental characteristic observed in numerous stars across the optical band, manifesting as periodic, semi-regular, or completely irregular brightness fluctuations. Variable stars are broadly categorized into \textit{intrinsic} and \textit{extrinsic} variables based on the underlying mechanisms driving their luminosity variations. 

\begin{table}[htbp]
    \centering
    \caption{Variable star types and their corresponding abbreviations~\cite{yu190}.}
    \label{table3}
    \begin{tabular}{ll}
    \toprule
    \textbf{Variable star types} & \textbf{Abbreviation} \\
    \midrule
    Eclipsing binary: Algol type & EA \\
    Eclipsing binary: Beta type & EB \\
    Eclipsing binary: W Ursae Majoris type & EW \\
    Ellipsoidal binaries & ELL \\
    Long period variable & LPV \\
    Mira & MIRA \\
    RV Tauri & RV \\
    W Virginis: period $<$8 d & CWA \\
    W Virginis: period $>$8 d & CWB \\
    RS Canum Venaticorum & RS \\
    BY Draconis & BY \\
    Population II Cepheid & PTCEPH \\
    Delta Cepheid & DCEP \\
    Delta Scuti & DSCT \\
    Gamma Doradus & GDOR \\
    B emission-line star & BE \\
    Gamma Cassiopeiae & GCAS \\
    Alpha Cygni & ACYG \\
    Beta Cephei & BCEP \\
    Alpha-2 Canum Venaticorum & ACV \\
    RR Lyrae: RRab type & RRAB \\
    RR Lyrae: RRC type & RRC \\
    RR Lyrae: RRd type & RRD \\
    Slowly pulsating B star & SPB \\
    \bottomrule
    \end{tabular}
\end{table}

Intrinsic variables experience genuine changes in luminosity due to internal physical processes, such as stellar pulsations, eruptions, or structural expansion and contraction. This category primarily includes pulsating and eruptive variables. Pulsating variables undergo periodic expansions and contractions in their outer layers, leading to observable brightness oscillations. Notable examples include Cepheids, RR Lyrae, and Mira variables, each exhibiting distinct pulsation periods and amplitude variations. Eruptive variables, such as cataclysmic variables and nova-like stars, exhibit sudden and often dramatic changes in brightness, typically due to stellar outbursts or accretion-related instabilities.

\begin{figure*}[htbp]
    \centering
    \includegraphics[width=0.95\textwidth]{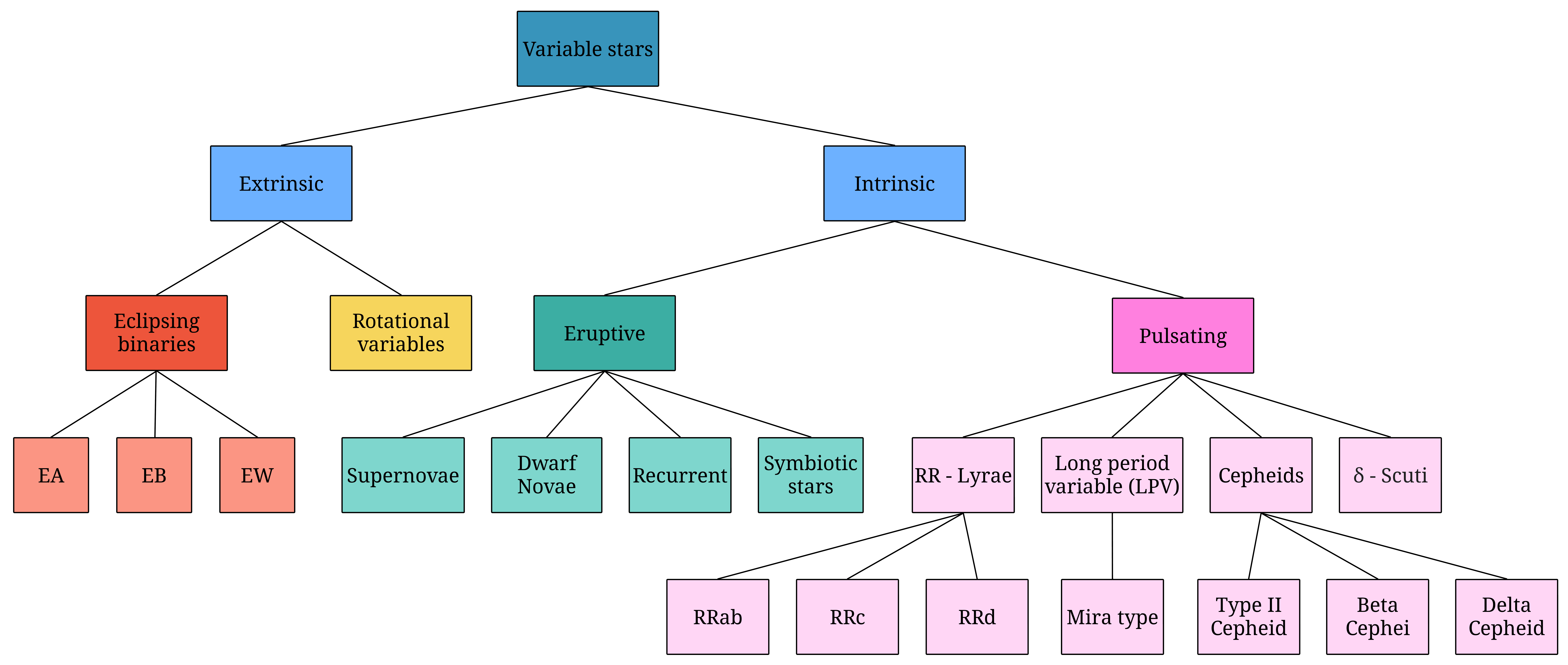}
    \caption{Variable star classification hierarchy~\cite{yu190}.}
    \label{fig:VS}
\end{figure*}

Extrinsic variables, in contrast, exhibit brightness fluctuations due to external factors, such as eclipses or rotational modulations. This category encompasses eclipsing binaries and rotating variables. Eclipsing binary systems consist of two or more gravitationally bound stars orbiting a common center of mass, where periodic eclipses result in characteristic minima in their LCs. The primary and secondary minima in their LCs provide insights into stellar radii, temperatures, and orbital inclinations. Rotating variables exhibit modest brightness variations arising from stellar surface features, such as starspots or ellipsoidal distortions, modulated by the star’s rotation.

Figure~\ref{fig:VS} presents a schematic classification of variable stars, while Table~\ref{table3} lists their primary types and corresponding abbreviations. The analysis of variable star LCs plays a crucial role in distinguishing between different classes. Pulsating variables, for instance, exhibit smooth and regular brightness oscillations, whereas eclipsing binaries show well-defined periodic minima. These distinctions enable robust classification and facilitate astrophysical inferences about stellar structure, evolution, and binary interactions.

The systematic classification of variable stars relies heavily on distinguishing subtle features in their LCs, such as periodicity, amplitude, and morphological patterns. Traditional classification frameworks often depend on manually engineered features (e.g., periodograms, Fourier coefficients, or phased-folded curve statistics), which may fail to capture nuanced or non-linear relationships in large datasets. ML methods overcome these limitations by automating feature extraction and enabling robust classification across diverse variable star populations. Table~\ref{table4} provides a comprehensive summary of ML techniques applied to variable star analysis, reflecting the evolution of methodologies from early Bayesian approaches to modern DL architectures.

Early ML implementations focused on probabilistic methods and ensemble techniques. Bayesian Networks (BNs) and Gaussian Mixture Models (GMMs) were employed for CoRoT and Hipparcos datasets to probabilistically associate LC features with physical classes~\cite{debosscher193,sarro192}. The introduction of RFs marked a significant advancement, enabling feature importance analysis and improved handling of imbalanced datasets~\cite{richards196,dubath197}. Subsequent hybrid approaches combined RFs with dimensionality reduction techniques like PCA or SOMs to enhance interpretability~\cite{armstrong141,rimoldini198}.

The advent of DL revolutionized variable star classification by leveraging raw or minimally preprocessed LCs. CNNs achieved state-of-the-art performance on surveys like CRTS and Kepler, identifying hierarchical patterns directly from flux measurements~\cite{mahabal211,akhmetali244}. RNNs and LSTM networks proved particularly effective for capturing temporal dependencies in irregularly sampled data from ASAS-SN and \textit{Gaia}~\cite{naul76,merino236}. Transformer architectures, recently applied to Kepler and ZTF data, demonstrated superior performance in modeling long-range dependencies without phase-folding assumptions~\cite{pan243,cadiz242}.

\begin{longtable*}{|m{3cm}|m{4cm}|m{10cm}|}
    \caption{Summary of ML methods applied to variable star analysis.}
    \label{table4} \\
    \hline
    \textbf{Data type} & \textbf{Method} & \textbf{Description} \\ \hline 
    \endfirsthead
    
    \multicolumn{3}{c}{{\tablename\ \thetable{} - Continued}} \\
    \hline
    \textbf{Data type} & \textbf{Method} & \textbf{Description} \\ \hline 
    \endhead
    
    \hline
    \endfoot
    
    \hline
    \endlastfoot
    
    ASAS & Bayesian Classifier &~\cite{eyer191} developed a Fourier-based   Bayesian classifier for ASAS variables. \\ \hline
    Hipparcos & Bayesian ensemble &~\cite{sarro192} developed a Bayesian ensemble   of neural networks for automatic classification of   eclipsing binary LCs. \\ \hline
    OGLE & BN and SVM &~\cite{debosscher193} presented BN and SVM   in the application of the methodology to variable stars. \\ \hline
    OGLE & BN and SVM &~\cite{sarro194} developed and tested a Fourier-based   Bayesian classifier \\ \hline
    CoRoT & BN and GMM &~\cite{debosscher195} developed a fast pipeline for   classifying CoRoT LCs and discovering new stellar   variability types. \\ \hline
    OGLE and Hipparcos & RF &~\cite{richards196} developed a ML methodology   using RF. \\ \hline
    Hipparcos & RF and BN &~\cite{dubath197} evaluated automated classification   of Hipparcos periodic stars using RFs. \\ \hline
    Hipparcos & RF and BN &~\cite{rimoldini198} applied RFs to classify periodic,   non-periodic, and irregular Hipparcos variables. \\ \hline
    Kepler & RF &~\cite{long199} introduced noisification to reduce   survey-dependent feature mismatch. \\ \hline
    Hipparcos and OGLE & Active Learning (AL) &~\cite{richards200} used active learning to reduce   sample bias in variable star classification \\ \hline
    PTF & RF &~\cite{bloom66}  developed a ML framework for PTF   to automate discovery and classification of transients   and variables. \\ \hline
    LINEAR & GMM and SVM &~\cite{sesar201} proposed method for identifying visually   confirmed variable stars within the LINEAR survey. \\ \hline
    LINEAR & GMM and SVM &~\cite{palaversa202} identified $\sim$7000 faint periodic stars   using LINEAR and SDSS data. \\ \hline
    MACHO & BN and RF &~\cite{nun203} proposed methodology for anomaly   detection in MACHO data. \\ \hline
    WISE & RF and AL &~\cite{masci64} proposed methodology for classifying    periodic variable stars. \\ \hline
    Catalogue of Eclipsing Variables & Membership probability &~\cite{avvakumova204} developed a procedure   to classify eclipsing binaries based on LC parameters. \\ \hline
    OGLE and ASAS & KNN, SVM and RF &~\cite{kugler205} proposed a density model for classifying   irregular time-series data. \\ \hline
    Kepler & SOM and RF &~\cite{armstrong141} developed a novel method for   classifying variable stars by combining SOM and RF. \\ \hline
    Kepler & RF and BN &~\cite{bass206} proposed an ensemble approach for   variable star classification in the Kepler field. \\ \hline
    MACHO, LINEAR and ASAS & RF &~\cite{kim207} developed a general-purpose   ML package for classifying periodic variable stars. \\ \hline
    MACHO and OGLE & SVM &~\cite{mackenzie208} extracted subsequences of LCs   and clustered them to identify common local patterns. \\ \hline
    UCR and LINEAR & KNN, RF, RBF-NN &~\cite{johnston209} developed a novel time-domain   feature extraction method called Slotted Symbolic Markov   Modeling (SSMM). \\ \hline
    ASAS, Hipparcos and OGLE & KNN, SVM, and RF &~\cite{johnston210} proposed a method for classifying   variable stars using supervised pattern recognition. \\ \hline
    CRTS & CNN &~\cite{mahabal211} developed a DL approach for   classifying LCs by transforming sparse, irregular time-series   data into 2D representation. \\ \hline
    ASAS, LINEAR, MACHO & RNN &~\cite{naul76} developed an unsupervised autoencoding   RNN that effectively handles irregularly sampled, noisy LCs. \\ \hline
    Kepler & LSTM and RNN &~\cite{hinners212} presented methods for representation   learning and feature engineering aimed at predicting and   classifying properties. \\ \hline
    OGLE & LR, SVM, KNN, RF, and SGB &~\cite{pashchenko120} proposed a ML approach for   variability detection. \\ \hline
    OGLE, MACHO and Kepler & Fast Similarity Function &~\cite{valenzuela213} presented a novel data   structure called the Variability Tree. \\ \hline
    \textit{Gaia} and ASAS & RF &~\cite{jayasinghe214} utilized the RF classifier along   with a series of classification corrections. \\ \hline
    OGLE, VISTA and CoRoT & CNN &~\cite{aguirre215} developed a scalable CNN architecture   for survey-independent LC classification. \\ \hline
    ASAS-SN & RNN &~\cite{tsang216} developed classifier combining an   RNN AE with a GMM. \\ \hline
    ASAS & PCA and RF &~\cite{mcwhirter217} focused on processing time-series   data with uneven cadence by leveraging representation   learning to extract useful features. \\ \hline
    CRTS & RF &~\cite{hosenie218} developed an optimized ML framework   for variable star classification. \\ \hline
    Light curves in Galactic Plane & $k$-medoids method &~\cite{modak219} proposed a $k$-medoids clustering   approach to objectively classify galactic variable stars. \\ \hline
    CoRoT, OGLE and MACHO & Streaming Probabilistic Model &~\cite{zorich220} proposed a streaming probabilistic   classification model that uses a novel set of features. \\ \hline
    OGLE, \textit{Gaia} and WISE & RNN &~\cite{becker221}  developed an end-to-end DL approach   using RNNs for efficient variable star classification. \\ \hline
    CRTS & RF and XGBoost &~\cite{hosenie222} proposed a hybrid approach combining   HC with data augmentation techniques. \\ \hline
    MACHO & RNN &~\cite{jamal223} conducted systematic comparison of   neural network architectures for time-series classification. \\ \hline
    UCR Starlight and LINEAR & Multi-View Metric Learning &~\cite{johnston224} introduced a  Multi-View Metric    Learning framework that leverages multiple data   representations. \\ \hline
    OGLE & iTCN and iResNet &~\cite{zhang225} developed Cyclic-Permutation Invariant   Neural Networks that achieve state-of-the-art accuracy. \\ \hline
    OGLE & CNN and LSTM &~\cite{bassi226} proposed 1D CNN-LSTM hybrid network   for direct variable star classification using raw time-series data. \\ \hline
    Simulated data & LSTM &~\cite{vcokina227} developed a DL method for automated   classification of eclipsing binaries. \\ \hline
    ZTF & BRF &~\cite{sanchez228} introduced ALeRCE's first LC   classifier, a two-level balanced RF system processing ZTF   alerts. \\ \hline
    Kepler & GMM &~\cite{barbara229}  developed an interpretable classification   system for Kepler LCs. \\ \hline
    OGLE & Multiple-Input Neural Network  &~\cite{szklenar230} developed a Multiple-Input Neural   Network combining CNNs and MLPs. \\ \hline
    OGLE, CSS, \textit{Gaia} & UMAP and HDBSCAN  &~\cite{pantoja231} developed semi-supervised and   clustering-based approaches for variable star classification.  \\ \hline
    VVV & RF and XGBoost &~\cite{molnar232} developed \texttt{VIVACE}, an automated    two-stage classification pipeline. \\ \hline
    ZTF & CVAE &~\cite{chan233} proposed an unsupervised DL approach using variational AE and isolation forests. \\ \hline
    Kepler & ResNet and LSTM &~\cite{yan234} developed RLNet, a hybrid ResNet-LSTM   neural network. \\ \hline
    TESS & SVM &~\cite{elizabethson235} developed a ML framework   classifying  T Tauri stars into 11 morphological classes. \\ \hline
    \textit{Gaia} & LSTM and GRU &~\cite{merino236} proposed a self-supervised learning   approach using RNNs. \\ \hline
    LAMOST & LightGBM and XGBoost &~\cite{qiao237} developed a LightGBM/XGBoost-based   classification system for LAMOST DR9 data. \\ \hline
    OGLE & CNN &~\cite{monsalves238} developed an efficient CNN-based   classification system using 2D histogram representations   of OGLE LCs. \\ \hline
    TESS & CNN &~\cite{olmschenk239} developed a rapid CNN classifier for   TESS 30-minute cadence data. \\ \hline
    TMTS & XGBoost and RF &~\cite{guo240} developed a classification system for TMTS   variables using XGBoost and RF. \\ \hline
    ZTF & Distance Metric Classifier &~\cite{chaini241} developed \texttt{DistClassiPy}, an interpretable   distance-metric classifier for variable stars. \\ \hline
    MACHO, OGLE and ATLAS & Transformer &~\cite{cadiz242} proposed HA-MC Dropout, a novel   transformer-based method combining hierarchical attention    and Monte Carlo dropout. \\ \hline
    Kepler & Transformer &~\cite{pan243} developed \texttt{Astroconformer}, a Transformer- based model that demonstrates superior performance. \\ \hline
    OGLE & CNN &~\cite{akhmetali244} developed a CNN-based approach for   automated variable star classification. \\ \hline
    
\end{longtable*}

\subsection{Supernova classification}\label{sec:Supernova}

Supernovae (SNe) are among the most energetic transient phenomena in the universe. They play a critical role in stellar evolution, the chemical enrichment of the interstellar medium, and cosmological distance measurements. Their classification traditionally relies on spectroscopic and photometric observations. While spectroscopic classification remains the most accurate, it is expensive and has high requirements for telescopes and observation time, and thus cannot be applied to all observed transients.

Photometric classification, although less precise, offers higher observational efficiency and has gained importance with the advent of wide-field surveys. Early photometric methods used template fitting and parametric modeling of LCs, leveraging features such as peak brightness, decline rate, and color evolution. However, these methods typically require complete LCs with full phase coverage, limiting their application to real-time or sparsely sampled data.

Recent advances in ML have significantly enhanced SN classification capabilities, especially under constraints such as low signal-to-noise ratios and incomplete data. ML-based approaches can classify a wide range of transient types and support near real-time classification. This is critical for follow-up prioritization and maximizing the scientific return from transient surveys. Table~\ref{table5} provides SN classes and their physical origins, while Table~\ref{table6} summarizes various ML methods applied to SN classification.

\begin{table}[htbp]
    \centering
    \caption{Major supernova classes and their physical origin~\cite{gal273}.}
    \label{table5}
    \begin{tabular}{ll}
    \toprule
    \textbf{Supernova types} & \textbf{Physical origin} \\
    \midrule
    SN Ia & White Dwarf \\
    SN Ib & Massive star \\
    SN Ic & Massive star \\
    SN Ic-BL & Massive star \\
    SN II & Massive star \\
    SN IIb & Massive star \\
    SN IIn & Massive star \\
    SLSN & Massive star \\
    \bottomrule
    \end{tabular}
\end{table}

Early ML applications focused on engineered features derived from parametric LC fits or domain-specific metrics. The Supernova Photometric Classification Challenge (SNPCC)~\cite{kessler272} was created as a standardized dataset to evaluate ML-based photometric classification methods for SN. Many subsequent studies tested their algorithms on the SNPCC dataset~\cite{newling245, richards62, karpenka246, gupta247, lochner73, charnock249, ishida77, pasquet254, santos258, de266}.

DL revolutionized SN classification by allowing models to process raw flux measurements or minimally preprocessed LCs. Initial ML applications employed a variety of feature‐based techniques: Kernel Density Estimation (KDE) and Boosting methods~\cite{newling245}, Non‐linear Dimensionality Reduction with RFs~\cite{richards62}, ANNs~\cite{karpenka246}, Domain Adaptation with AL~\cite{gupta247}, and Naive Bayes, KNN, SVM, ANN, and BDTs~\cite{lochner73}. 

RNN and CNN architectures then advanced photometric classification by ingesting raw time series directly. Deep RNNs (including LSTM variants) demonstrated strong performance on SNPCC and SALT2‐fitted LCs~\cite{charnock249,moller255}, while CNNs learned hierarchical features from 2D representations of LCs~\cite{brunel250,qu262}. Hybrid models, such as \texttt{SuperNNova}~\cite{moller255}, \texttt{SuperRAENN}~\cite{villar256}, and \texttt{PELICAN} framework~\cite{pasquet254} leveraged SSL and AEs to boost purity and completeness.

Recent classification pipelines increasingly incorporate generative models, Gaussian Process (GP) augmentation, and real-time alert systems~\cite{Martínez-Palomera_2022, boone261, smith53}. For example, the \texttt{ParSNIP} framework~\cite{boone261} utilizes Variational Autoencoders (VAEs) to perform classification. Similarly, \texttt{Avocado}~\cite{boone253} applies LightGBM combined with GP augmentation to classify transients photometrically. Real-time alert brokers like \texttt{Fink}~\cite{leoni264} streamline the early identification of SNe from surveys like ZTF using AL strategies. CNN-based models such as \texttt{SCONE}~\cite{qu262} apply 2D GP regression to multi-band LCs, while \texttt{Photo-$z$SNthesis}~\cite{qu268} uses CNNs to generate full redshift probability distributions. Temporal convolutional networks (TCNs) paired with LightGBM, as in the \texttt{TLW} model~\cite{li271}, further demonstrate the effectiveness of hybrid architectures for robust, survey-independent transient classification.

\begin{longtable*}{|m{3cm}|m{5cm}|m{9cm}|}
    \caption{Summary of ML methods applied to supernova classification.}
    \label{table6} \\
    \hline
    \textbf{Data type} & \textbf{Method} & \textbf{Description} \\ \hline 
    \endfirsthead
    
    \multicolumn{3}{c}{{\tablename\ \thetable{} - Continued}} \\
    \hline
    \textbf{Data type} & \textbf{Method} & \textbf{Description} \\ \hline 
    \endhead
    
    \hline
    \endfoot
    
    \hline
    \endlastfoot
    
    SNPCC & KDE and Boosting &~\cite{newling245} proposed two classification methods for the application of SNPCC data. \\ \hline
    SNPCC & Non-linear Dimension Reduction and RF &~\cite{richards62} proposed the non-linear dimension reduction technique  to detect structure in a data base of SNe LCs.  \\ \hline
    SNPCC & ANN &~\cite{karpenka246} presented a method for automated photometric classification of SNe.  \\ \hline
    SNPCC & Domain Adaptation and AL &~\cite{gupta247} presented an adaptive mechanism that generates  a predictive model to identify SNe Ia.  \\ \hline
    SNPCC & Naive Bayesian, KNN, SVM, ANN and BDT &~\cite{lochner73} developed a multi-faceted classification pipeline.  \\ \hline
    SNLS & XGBoost &~\cite{moller248} presented a method to photometrically classify SNe Ia. \\ \hline
    SNPCC & Deep RNN &~\cite{charnock249} presented deep RNN for performing photometric classification of SNe. \\ \hline
    SNPCC & CNN &~\cite{brunel250} presented CNN for SNe Ia classification. \\ \hline
    SNPCC & AL &~\cite{ishida77} developed a framework for spectroscopic follow-up design for optimizing supernova photometric classification. \\ \hline
    PS1-MDS & SVM, RF and MLP &~\cite{villar251} developed 24 classification pipelines with different feature extraction and data augmentation methods. \\ \hline
    PLAsTiCC & DNN &~\cite{muthukrishna252} developed \texttt{RAPID}, a novel time-series classication tool for identifying explosive transients.  \\ \hline
    PLAsTiCC & LightGBM &~\cite{boone253} developed \texttt{Avocado}, a software package for classification of transients with GP augmentation. \\ \hline
    SNPCC & CNN and AE &~\cite{pasquet254} developed \texttt{PELICAN}, an algorithm for the characterization and the classification of SNe LCs. \\ \hline
    SALT2 fitted & RNN &~\cite{moller255} developed \texttt{SuperNNova}, a framework for photometric classification of SNe. \\ \hline
    PS1-MDS & RF and RAENN &~\cite{villar256} developed \texttt{SuperRAENN}, a semi-supervised SN photometric classification pipeline. \\ \hline
    OSC and ZTF & RF &~\cite{gomez257} developed a classification algorithm targeted at rapid identification of a pure sample of SLSN-I. \\ \hline
    SNPCC & TPOT, XGBoost, AdaBoost, GBoost, EXT, RF &~\cite{santos258} analyzed the performance of boosting and averaging methods for classification of SNe. \\ \hline
    SALT2 & DNN &~\cite{takahashi259} developed a classification algorithm to classify LCs observed by Subaru/HSC. \\ \hline
    PS1-MDS & RF &~\cite{hosseinzadeh260} developed \texttt{Superphot}, an open-source classification algorithm for photometric classification of SNe. \\ \hline
    PS1-MDS and PLAsTiCC & VAE &~\cite{boone261} developed \texttt{ParSNIP}, a hybrid model to produce empirical generative models of transients from data sets of unlabeled LCs. \\ \hline
    PLAsTiCC & CNN &~\cite{qu262} developed \texttt{SCONE}, a CNN-based classification method using 2D GP regression. \\ \hline
    PLAsTiCC & CNN &~\cite{qu263} presented classification results on early SNe LCs from \texttt{SCONE}. \\ \hline
    ZTF & AL &~\cite{leoni264} developed \texttt{Fink}, a broker for early SNe classification. \\ \hline
    Open Supernova Catalog & GP &~\cite{stevance265} explored the application of GP to SNe LCs. \\ \hline
    SNPCC & XGBoost &~\cite{de266} developed a linear regression algorithm optimized through automated machine learning (AutoML) frameworks. \\ \hline
    PLAsTiCC and ZTF & MLP, NF and Bayesian Neural Network &~\cite{demianenko267} examined several ML-based LC approximation methods. \\ \hline
    PLAsTiCC and Simulated SDSS-II SN data & CNN &~\cite{qu268} developed \texttt{Photo-$z$SNthesis}, a CNN-based method for predicting full redshift probability distributions from multi-band SNe LCs. \\ \hline
    ZTF & AL &~\cite{pruzhinskaya269} explored the potential of AL techniques in application to detect new SNe candidates. \\ \hline
    ZTF & LightGBM &~\cite{de270} developed \texttt{Superphot+}, a photometric classifier for SNe LCs that does not rely on redshift information. \\ \hline
    PLAsTiCC & TCN and LightGBM &~\cite{li271} developed \texttt{TLW}, a classification algorithm for transients. \\ \hline

\end{longtable*}

\section{Challenges and Open Issues}\label{sec:Challenges}

The application of ML to LC analysis has revolutionized astronomical research, yet significant challenges remain in developing robust, interpretable, and survey-independent classification systems. These challenges are expected to intensify with upcoming data-intensive surveys such as the \textit{LSST}, which will detect tens of millions of transient events nightly.

A key difficulty lies in the heterogeneity and sparsity of data. Surveys like \textit{ASAS-SN} and \textit{Gaia} produce irregularly sampled LCs due to observational constraints such as weather or cadence gaps. This irregularity complicates the use of DL models, which typically assume regularly sampled inputs. Moreover, models trained on one survey (e.g., \textit{Kepler}) often perform poorly when applied to others (e.g., \textit{TESS}), due to differences in cadence, noise levels, and photometric filters~\cite{valizadegan178}. These issues are further exacerbated by limited spectroscopic follow-up, especially for rare classes such as SLSNe.

Model interpretability and physical consistency are additional concerns. Many ML models act as black boxes, offering little insight into their predictions. This lack of transparency hinders their adoption in cosmological studies that require rigorous uncertainty quantification and interpretability. Hybrid approaches that incorporate physical priors into data-driven models are a promising direction.

Real-time processing and early classification present another critical challenge. \textit{LSST} will require sub-minute response times for alert streams, straining even highly optimized models. Furthermore, most classifiers rely on full-phase LCs, whereas early-time classification should operate with incomplete data—often limited to the rising or plateau phases. Building models that can deliver accurate early-stage predictions is an active area of research.

Class imbalance and anomaly detection remain persistent issues. Rare transients such as kilonovae and luminous red novae are difficult to detect using conventional techniques. Oversampling strategies often fall short, prompting interest in generative and data augmentation methods. Moreover, most classifiers assume a closed-set of known classes, yet upcoming surveys will likely discover entirely new types of transients. 

Generative models—such as GANs, VAEs, and diffusion models—are emerging as powerful tools for addressing these challenges. They enable synthetic augmentation of rare classes, imputation of missing or irregular data, and cross-survey domain adaptation. While still in early development, generative approaches show strong potential to improve the robustness and generalizability of ML-based LC pipelines.

\section{Conclusions}\label{sec:Conclusions}

The rapid advancement of observational capabilities in astronomy has led to an exponential increase in the volume of LC data, opening up both exciting opportunities and complex challenges for time-domain astronomy. In this evolving landscape, ML has emerged as a powerful tool, finding applications across a broad spectrum of tasks. This review discusses major photometric surveys that provide the essential LC data, outlines the fundamental principles of ML, and explores ML applications in LC analysis, including exoplanet detection, variable star analysis, and supernova classification, highlighting the increasing sophistication and versatility of these methods.

As astronomical surveys scale up in depth, cadence, and volume, the need for automated, scalable, and interpretable analysis pipelines becomes ever more urgent. ML models, particularly DL architectures, have shown exceptional performance in handling large, noisy, and irregular datasets. Importantly, the choice of ML approach depends on the specific scientific goal, the nature of the dataset, and the balance among performance, interpretability, and computational cost. At the same time, critical challenges such as survey dependence, class imbalance, interpretability, and real-time applicability remain open issues.

Future advancements are likely to involve approaches that go beyond purely data-driven methods by integrating physical models, enhancing generalizability across different surveys, and incorporating robust uncertainty quantification.The convergence of domain expertise and machine intelligence holds the key to not only improving classification accuracy but also enabling new scientific insights.

In conclusion, the fusion of ML with astronomical time-series data is not just a technical advancement—it represents a paradigm shift in how discoveries are made. As datasets continue to expand, so too will the opportunities for ML to illuminate the dynamic universe in ways previously unimaginable.

\bibliography{manuscipt}{}
\bibliographystyle{aasjournal}

\end{document}